\documentclass[acmsmall,screen,authorversion,nonacm]{acmart}

\usepackage{xcolor,colortbl}
\usepackage{graphicx}
\usepackage{color}
\usepackage{pifont}
\usepackage{ifthen}
\usepackage{multirow}
\usepackage{booktabs}
\usepackage[inline]{enumitem}
\usepackage[ruled,linesnumbered]{algorithm2e}
\usepackage{tcolorbox}
\usepackage{booktabs}
\usepackage{tabularx}
\newcolumntype{L}{>{\raggedright\arraybackslash}X}%
\newcolumntype{R}{>{\raggedleft\arraybackslash}X}%
\newcolumntype{C}{>{\centering\arraybackslash}X}%
\usepackage{listings}
\usepackage{relsize}
\usepackage{threeparttable}
\usepackage{amsmath}
\usepackage{mathtools}
\usepackage{tikz}
\usepackage{fontawesome}
\usepackage{caption}
\usepackage{subcaption}
\usepackage{titlesec}
\usepackage{xspace}

\usepackage{gnuplottex}

\usetikzlibrary{shapes,positioning}
\makeatletter
\let\old@lstKV@SwitchCases\lstKV@SwitchCases
\def\lstKV@SwitchCases#1#2#3{}
\makeatother
\usepackage{lstlinebgrd}
\makeatletter
\let\lstKV@SwitchCases\old@lstKV@SwitchCases

\lst@Key{numbers}{none}{%
    \def\lst@PlaceNumber{\lst@linebgrd}%
    \lstKV@SwitchCases{#1}%
    {none:\\%
        left:\def\lst@PlaceNumber{\llap{\normalfont
                \lst@numberstyle{\thelstnumber}\kern\lst@numbersep}\lst@linebgrd}\\%
        right:\def\lst@PlaceNumber{\rlap{\normalfont
                \kern\linewidth \kern\lst@numbersep
                \lst@numberstyle{\thelstnumber}}\lst@linebgrd}%
    }{\PackageError{Listings}{Numbers #1 unknown}\@ehc}}
\makeatother

\def\editmode{}

\ifthenelse{\isundefined{\editmode}}{
\newcommand{\editnote}[3]{%
}

}{
\newcommand{\editnote}[3]{\xspace\colorbox{#1}{\sffamily \smaller \textcolor{white}{~\faCommenting{}~#2~}}\textcolor{#1}{~#3}\xspace}

}

\definecolor{nord0}{HTML}{2E3440}
\definecolor{nord1}{HTML}{3B4252}
\definecolor{nord2}{HTML}{434C5E}
\definecolor{nord3}{HTML}{4C566A}
\definecolor{nord4}{HTML}{D8DEE9}
\definecolor{nord5}{HTML}{E5E9F0}
\definecolor{nord6}{HTML}{ECEFF4}
\definecolor{nord7}{HTML}{8FBCBB}
\definecolor{nord8}{HTML}{88C0D0}
\definecolor{nord9}{HTML}{81A1C1}
\definecolor{nord10}{HTML}{5E81AC}
\definecolor{nord11}{HTML}{BF616A}
\definecolor{nord12}{HTML}{D08770}
\definecolor{nord13}{HTML}{EBCB8B}
\definecolor{nord14}{HTML}{A3BE8C}
\definecolor{nord15}{HTML}{B48EAD}

\newcommand{\tool}{\textsc{StubCoder}\@\xspace}
\newcommand{\mockito}{\textsc{Mockito}\@\xspace}
\newcommand{\easymock}{\textsc{EasyMock}\@\xspace}
\newcommand{\java}{\textsc{Java}\@\xspace}
\newcommand{\moq}{\textsc{Moq4}\@\xspace}
\newcommand{\junit}{\textsc{JUnit}\@\xspace}
\newcommand{\arja}{\textsc{Arja}\@\xspace}
\newcommand{\cardumen}{\textsc{Cardumen}\@\xspace}
\newcommand{\astor}{\textsc{Astor}\@\xspace}
\newcommand{\eg}[1]{(e.g., #1)}
\newcommand{\ie}[1]{(i.e., #1)}
\newcommand{\etal}[0]{et al.}

\newcommand{\numOfBenchmarkProjects}{13}
\newcommand{\numOfBenchmarkEntries}{59}
\newcommand{\testTuple}{\tau{} = \left\langle{} V, S, E, A\right\rangle}

\newcommand{\code}[1]{\text{\ttfamily#1}}

\definecolor{boxbg}{RGB}{240, 240 ,240}

\newcounter{rq}
\newenvironment{answertorq}{%
    \begin{tcolorbox}[
            arc=2mm,
            boxrule=0.5pt,
            left=2pt,
            right=2pt,
            top=2pt,
            bottom=2pt
        ]
    \textbf{\faLightbulbO\ RQ\refstepcounter{rq}\therq{} in Summary:}
}{\end{tcolorbox}}

\tcbuselibrary{breakable}

\titleformat*{\paragraph}{\bfseries}

\setcopyright{cc}
\setcctype{by-nc-nd}

\begin{document}

\title{\tool: Automated Generation and Repair of Stub Code for Mock Objects}
\titlenote{This paper was accepted by the \emph{ACM Transactions on Software Engineering and Methodology} (TOSEM) in July 2023.}

\acmJournal{TOSEM}



\author{Hengcheng Zhu}
\orcid{0000-0002-3082-5957}
\email{hzhuaq@connect.ust.hk}
\affiliation{%
    \institution{The Hong Kong University of Science and Technology}
    \city{Hong Kong}
    \country{China}
}

\author{Lili Wei}
\orcid{0000-0002-2428-4111}
\email{lili.wei@mcgill.ca}
\affiliation{%
    \institution{McGill University}
    \city{Montreal}
    \country{Canada}
}

\author{Valerio Terragni}
\orcid{0000-0001-5885-9297}
\email{v.terragni@auckland.ac.nz}
\affiliation{%
    \institution{The University of Auckland}
    \city{Auckland}
    \country{New Zealand}
}

\author{Yepang Liu}
\orcid{0000-0001-8147-8126}
\email{liuyp1@sustech.edu.cn}
\affiliation{%
    \department{Department of Computer Science and Engineering}
    \department{Research Institute of Trustworthy Autonomous Systems}
    \institution{Southern University of Science and Technology}
    \city{Shenzhen}
    \country{China}
}

\author{Shing-Chi Cheung}
\authornote{Shing-Chi Cheung is the corresponding author.}
\orcid{0000-0002-3508-7172}
\email{scc@cse.ust.hk}
\affiliation{%
    \institution{The Hong Kong University of Science and Technology}
    \city{Hong Kong}
    \country{China}
}

\author{Jiarong Wu}
\orcid{0000-0001-6126-303X}
\email{jwubf@cse.ust.hk}
\affiliation{%
    \institution{The Hong Kong University of Science and Technology}
    \city{Hong Kong}
    \country{China}
}

\author{Qin Sheng}
\orcid{0009-0004-8527-9297}
\email{entersheng@webank.com}

\author{Bing Zhang}
\orcid{0009-0007-5862-136X}
\email{billzzhang@webank.com}

\author{Lihong Song}
\orcid{0009-0004-9452-4741}
\email{jeaninesong@webank.com}
\affiliation{%
    \institution{WeBank Co. Ltd.}
    \city{Shenzhen}
    \country{China}
}

\renewcommand{\shortauthors}{H. Zhu et al.}

\received{August 10, 2022}
\received[revised]{April 11, 2023}
\received[revised]{June 15, 2023}
\received[accepted]{July 23, 2023}

\begin{abstract}
    Mocking is an essential unit testing technique for isolating the class under test (CUT) from its dependencies.
Developers often leverage mocking frameworks to develop stub code that specifies the behaviors of mock objects.
However, developing and maintaining stub code is labor-intensive and error-prone.
In this paper, we present \tool{} to automatically generate and repair stub code for regression testing.
\tool{} implements a novel evolutionary algorithm that synthesizes test-passing stub code guided by the runtime behavior of test cases. 
We evaluated our proposed approach on \numOfBenchmarkEntries{} test cases from \numOfBenchmarkProjects{} open-source projects.
Our evaluation results show that \tool{} can effectively generate stub code for incomplete test cases without stub code and repair obsolete test cases with broken stub code.
\end{abstract}

\maketitle

\section{Introduction}
Unit testing is an important testing paradigm that focuses on the correctness of a single software component~\eg{class in \textsc{Java}}~\cite{DBLP:books/daglib/0020331}.
In practice, a class is commonly implemented to leverage other classes' functionality.
These classes constitute \emph{test dependencies}, which are invoked when testing the \emph{class under test} (CUT).
To test the CUT in isolation, developers often substitute dependencies with  \emph{test doubles}~\cite{mcdonough2021test}, which play the role of dependencies for testing purpose only.
There are five main types of test doubles: dummy, stub, mock, spy, and fake~\cite{mcdonough2021test}.
Following the popular terminology of the \mockito{} framework~\cite{Tool:mockito}, we use the term \emph{mock objects}~\cite{DBLP:journals/ese/SpadiniABB19} to collectively refer to dummy, stub, and mock test doubles.
The mock objects in \mockito can act like any of these three types of test doubles in a unit test~\cite{Tool:mockito}.
In a nutshell, mock objects are designed to simulate the reactions of test dependencies~\ie{via stub calls} or validate their interactions with the CUT~\ie{via mocking calls}~\cite{thomas2002mock}.

\definecolor{stubbg}{HTML}{FCF3D5}
\begin{figure}[t]
\begin{lstlisting}[
		language=java,
		morekeywords={var},
		caption={An Illustration of Unit Test with Mock Objects},
		label={lst:stubbing-example},
		escapechar=|,
		linebackgroundcolor = {\ifnum \value{lstnumber} > 6 \ifnum \value{lstnumber} < 12 \color{stubbg} \fi \fi},
		numbers=left
	]
@Test
public void loginRetryTest() {
    // Arrange: Prepares the dependencies and input data
    //          for the class under test.
    var dao = mock(UserDao.class); // mock object|\label{code:stubbing-example:arrange-begin}|
    var user = mock(User.class);   // mock object|\label{code:stubbing-example:extra-mock}|
    var sha1 = DigestUtils.sha1Hex("bar"); |\label{code:stubbing-example:stubbing-begin}|         |STUB CODE|
    when(user.getPasswordHash()).thenReturn(sha1); // Stub Call 1|\label{code:stubbing-example:stubbing-dao}| 
    when(dao.findUser(eq("foo"))                   // Stub Call 2
        .thenThrow(new TimeoutException())         // First reaction |\label{code:stubbing-example:first-reaction}|
        .thenReturn(user);                         // Second reaction |\label{code:stubbing-example:arrange-end}|

    // Act: Exercises the feature
    var underTest = new LoginService(dao);|\label{code:stubbing-example:act-begin}|
    var loginResult = underTest.login("foo", "bar");|\label{code:stubbing-example:act-end}|

    // Assert: Verifies the result.
    verify(dao, times(2)).findUser(any());// mocking call|\label{code:stubbing-example:assert-begin}|
    assertTrue(loginResult.isSuccessful());|\label{code:stubbing-example:assert-end}|
}

// CUT
public LoginResult login(String userName, String password) {|\label{code:stubbing-example:cut-begin}|
    User user = null;
    while(user == null) {|\label{code:stubbing-example:loop}|
        try {
            user = dao.findUser(userName);
        } catch (TimeoutException e) { ... }|\label{code:stubbing-example:catch}|
    }
    var hash = DigestUtil.sha1Hex(password);
    if (hash.equals(user.getPasswordHash()))|\label{code:stubbing-example:hash-check}|
        // success
    else
        // failure
}|\label{code:stubbing-example:cut-end}|
\end{lstlisting}
\end{figure}

Listing~\ref{lst:stubbing-example} illustrates a \junit test case with mock objects implemented using the \mockito{} framework~\cite{Tool:mockito}.
The unit test aims to validate the login function of \code{LoginService}, which leverages its test dependency \code{UserDao} to establish a database connection.
The test case simulates the dependent database service \code{UserDao} and a database entity \code{User} with two mock objects.
Lines~\ref{code:stubbing-example:stubbing-begin}--\ref{code:stubbing-example:arrange-end} give the stub code that specifies the behaviors of the mock objects when their methods are (indirectly) invoked at Line~\ref{code:stubbing-example:act-end}.
The first invocation of the method \code{findUser} throws an exception to simulate an unstable connection (Line~\ref{code:stubbing-example:first-reaction}).
The second invocation of the method \code{findUser} returns a \code{User} object to simulate a successful database query (Line~\ref{code:stubbing-example:arrange-end}).
The returned \code{User} object is another mock object to simulate a database entity. It returns the SHA-1 digest of a predefined password when its method \code{getPasswordHash} is invoked (Line~\ref{code:stubbing-example:stubbing-begin}).
At Line~\ref{code:stubbing-example:assert-begin}, a mocking assertion \code{verify} checks whether the CUT invoked the  method \code{findUser} twice.

With such mock objects, developers no longer need to set up a database for testing or unplug the network cable to trigger an exception.
Similar to the example, developers often replace dependencies with mock objects and specify their behaviors with stub code when the dependencies are hard to set up, flaky, faulty, or even not yet implemented~\cite{DBLP:journals/ese/SpadiniABB19,DBLP:conf/kbse/ZhuWWLCSZ20}.
The use of mock objects allows developers to focus on the CUT without worrying about the correctness or availability of its dependencies.

Developing and maintaining stub code is challenging.
When developing stub code for a mock object, developers need to carefully consider its possible interactions with the CUT, and simulate the reactions accordingly when its methods are called.
Stub code is tightly coupled with a specific implementation of the CUT (and its dependencies) and is easy to get broken when the implementation of CUT evolves.
Take the test case in Listing~\ref{lst:stubbing-example} as an example.
When the implementation of \code{UserDao}, \code{User}, or \code{LoginService} changes, the stub code will become broken since it no longer specifies the behaviors for the APIs needed by the test case.
For example, when the signature of \code{findUser} is changed from \code{findUser(userName)} to \code{findUser(userName, passwordHash)}, the stub code are broken and does not compile.
In this case, the stub code needs to be updated to adapt to the new implementation.
In real-world projects, such updates need to be done frequently to keep the behaviors of mock objects consistent with the production code~\cite{DBLP:journals/ese/SpadiniABB19}.
This activity is labor-intensive and error-prone~\cite{DBLP:conf/icse/FazziniCCLKGO22}.

Previous works on automatic stub code generation for mock objects rely on a capture-and-replay approach~\cite{DBLP:conf/paste/SaffE04,DBLP:conf/kbse/SaffAPE05,DBLP:conf/kbse/FazziniGO20,DBLP:conf/icsm/JoshiO07,DBLP:conf/sigsoft/ElbaumCDD06}.
Given an executable test case, such techniques generate stub code in three steps: (1) execute the test cases capturing the interactions between the CUT and its dependencies, (2) replace the dependencies with mock objects, and (3) create stub code according to the captured interactions.
As such, capture-and-replay techniques are able to generate stub code for only those test cases without mock objects.
This is because they need to invoke the actual dependencies.
However, the study of Spadini \etal{}~\cite{DBLP:journals/ese/SpadiniABB19} shows that for 83\% of test cases that use mock objects, the mock objects are introduced when the actual dependencies are hard to set up, flaky, or unavailable.
Therefore, capture-and-replay techniques are inapplicable to the majority of situations where mock objects are used.

Our goal is to synthesize stub code for mock objects without capturing the actual behaviors of the dependencies. This is challenging because it requires identifying the desired mock object's behavior for a specific test case.
Indeed, mock objects are often test-specific because the same dependency class often has different behaviors in different test cases~\cite{DBLP:conf/kbse/ZhuWWLCSZ20}.

In regression testing, we want to synthesize stub code to test future versions of the CUT.
In such a context, we do not aim to generate a stub code that makes the test pass or fail depending on whether the current version is faulty or not.
We aim to generate the stub code that makes the test pass on the current CUT version, which aims to detect regression bugs introduced in future versions.
Our observation is that the expected behavior of such test-passing stub code is often encoded in the CUT execution code and test oracles.
For example, consider the test case of Listing~\ref{lst:stubbing-example} without the stub code
highlighted in yellow (Lines~\ref{code:stubbing-example:stubbing-begin} to~\ref{code:stubbing-example:arrange-end}).
The expected behavior of the test case is given by Lines~~\ref{code:stubbing-example:act-begin} and~\ref{code:stubbing-example:act-end}, which specify how the CUT should be invoked, and Lines~\ref{code:stubbing-example:assert-begin} and~\ref{code:stubbing-example:assert-end}, which specify the expected output of the method under test.
The desired stub code  (Lines~\ref{code:stubbing-example:stubbing-begin} to~\ref{code:stubbing-example:arrange-end}) is the one that makes the test pass.

In this paper, we present \textbf{\tool}, the first technique to automatically generate stub code without executing the actual dependencies.
Given an incomplete test case without stub code, \tool{} leverages the CUT execution code and test oracles as specifications to guide the synthesis of stub code to make the tests pass for the current implementation.
As mentioned above, the synthesized stub code satisfies the expected behavior of the test case in the regression testing scenario and could detect bugs in future versions.

Due to the huge search space of possible stub code, it is infeasible to randomly or systematically explore all the possible candidate stub code to find a test-passing one.
As such, we design \tool{} based on an evolutionary algorithm that drives the search by examining the runtime behavior of each candidate stub code.
In particular, \tool{} employs a novel fitness function that captures how close a candidate stub code is to test-passing stub code.
The fitness function captures several runtime aspects like the distance between the expected and actual value of each oracle assertion, which effectively directs the search towards the candidates that are more likely to pass the test.

Notably, \tool{} can also be used for repairing stub code that is broken due to code evolution.
In such cases, \tool{} prioritizes the selection of code elements in the broken stub code when constructing a new candidate stub code.
Indeed, test-passing stub code might be syntactically similar to the obsolete one.

\tool{} has two application scenarios: (1) When adding a new test case, developers can write the code that exercises the CUT using mock  objects and specify the expected behavior for the test case by writing oracle assertions \ie{\junit assertions and mocking calls}. \tool{} will automatically generate the stub code.
(2) When the stub code in some test cases becomes obsolete due to software evolution, developers can run \tool{} to repair the broken stub code.
By supporting these two scenarios, \tool{} helps relieve developers from the tedious manual effort of stub code development and maintenance.

We evaluated \tool{} on \numOfBenchmarkEntries{} test cases collected from \numOfBenchmarkProjects{} open-source projects in both application scenarios. Although modern program synthesis tools (e.g., GitHub Copilot) can suggest possible statements to complete test cases, they do not aim to synthesize  test-passing stub code.
Since there is no related tool that generates stub code under these scenarios, we compared with a variant of \tool{} based on an unguided strategy.
In the both scenario, \tool{} successfully generates stub code for 76\% of the test cases in half of the repetitions.
Compared with the unguided variant, \tool{} has a higher success rate and synthesizes the stub code faster.
Moreover, 57\% of the synthesized stub code have identical fault detection capability as those written by developers (measured by mutation coverage in Table~\ref{tab:fidelity-generation} and Table~\ref{tab:fidelity-repair}).

To summarize, this paper makes three major contributions:
\begin{itemize}
	\item
	      We design and implement \tool{}, the first automatic stub code synthesis technique that can effectively synthesize stub code for the test dependencies of a target unit test case.
	      We equipped \tool{} with a novel fitness function that examines the runtime behaviors of the test case to guide the search of the test-passing stub code.
	\item
	      We construct the first benchmark for evaluating stub code generation and repair techniques. It is composed of \numOfBenchmarkEntries{} test cases from \numOfBenchmarkProjects{} open-source projects.
	      \tool{} can effectively synthesize stub code for incomplete test cases in both application scenarios and it outperforms the baselines as well as its unguided variant.
	\item We publicly release \tool{} and the benchmark to facilitate future research in this area.
	      The dataset is available at \url{https://doi.org/10.5281/zenodo.7816758}.
\end{itemize}

The remainder of this paper is organized as follows:
Section~\ref{sec:problem-formulation} formulates the stub code synthesis problem with a motivating example and highlights the technical challenges.
Section~\ref{sec:approach} presents the design and implementation of \tool{}.
Section~\ref{sec:evaluation} describes our evaluation of \tool{} on 59 test cases collected from 13 open-source projects.
Section~\ref{sec:related-work} discusses the related work.
Section~\ref{sec:conclusion} concludes the paper and points out possible future work.

\section{Motivating Example \& Problem Formulation}\label{sec:problem-formulation}

In this section, we leverage the example in Listing~\ref{lst:stubbing-example} to illustrate how we formulate and address the problem of stub code generation and repair for unit tests.

\subsection{Formulation of Unit Test Cases}
Unit test cases are commonly executed in three phases, following the AAA pattern~\cite{DBLP:conf/icsm/YuTA19} \ie{Arrange, Act, Assert}.
First, the Arrange phase sets up the test environment, which includes the setup of dependencies with mock objects.
Next, the Act phase exercises the CUT by invoking its methods.
Finally, the Assert phase checks whether the CUT produces the expected test outputs.

We represent a test case as a tuple \( \tau = \langle V, S, E, A \rangle\).
Such representation is in line with the AAA pattern:
\begin{itemize}
	\item \textbf{Arrange Phase:}
	      \(V\) is the set of variables that are used in \(E\) and \(A\), and \(S\) is the stub code that specifies the behaviors of mock objects in \(V\).
	\item \textbf{Act Phase:}
	      \(E\) represents the bytecode instructions that exercise the CUT.
	\item \textbf{Assert Phase:}
	      \(A\) is the test oracle, including mocking calls.
\end{itemize}

Take the test case in Listing~\ref{lst:stubbing-example} as an example.
\(V\) contains two mock objects \code{dao} and \code{user} (Lines~\ref{code:stubbing-example:arrange-begin}--\ref{code:stubbing-example:extra-mock}).
\(S\) contains the stub code (the highlighted region) that sets up the behavior of the mock objects in \(M\).
For example, Lines~\ref{code:stubbing-example:stubbing-begin}--\ref{code:stubbing-example:stubbing-dao} set the behavior of \code{dao} and specify that its method \code{getPasswordHash} should return the SHA-1 digest of string \code{"bar"}.
\(E\) contains Lines~\ref{code:stubbing-example:act-begin}--\ref{code:stubbing-example:act-end} that exercise the login function of the CUT \code{LoginService}.
\(A\) contains Lines~\ref{code:stubbing-example:assert-begin}--\ref{code:stubbing-example:assert-end}, which check whether the login function performs as expected using a mocking call (Line~\ref{code:stubbing-example:assert-begin}) and a \junit{} assertion (Line~\ref{code:stubbing-example:assert-end}).

\subsection{Problem Statement}\label{ssec:problem-statement}
\begin{figure}[t]
\begin{lstlisting}[
		language=java,
		morekeywords={var},
		caption={Stub Code for Subject \#36},
		label={lst:benchmark-36rep},
		escapechar=|,
		numbers=left
	]
// Stub ode written by developers
when(endPoint.getPath("health")).thenReturn("/actuator/health");

// Stub code synthesized by StubCoder (Repair Mode)
String var0 = "/actuator/health";
doReturn(var0).when(pathMappedEndpoints).getPath(any());
\end{lstlisting}
\end{figure}
Following our formulation of unit test cases, we define our stub code synthesis problem in two application scenarios.

\paragraph{Scenario \#1: Generation Mode.}
Given an incomplete test case \(\tau = \left\langle V,\varnothing, E, A\right\rangle \) without stub code, generate \(S\) such that \(\testTuple \) passes~\ie{all the oracle assertions in \(A\) pass without uncaught exceptions}.

In this scenario, our technique helps developers to develop test cases that are independent of their test dependencies.
For example, in Listing~\ref{lst:stubbing-example}, we can synthesize the stub code in the highlighted region given the remaining lines such that the oracle assertions at Line~\ref{code:stubbing-example:assert-begin} and Line~\ref{code:stubbing-example:assert-end} hold.
When developers are creating a new test case, they can simply instantiate the CUT with mock dependencies and finish the remaining parts without needing to consider the possible interactions between the CUT and the dependencies.
After that, they can launch \tool{} to synthesize the stub code to complete the test case.
We target at regression testing in this scenario, where we assume that the current system is correct and try to capture regressions in future versions.

\paragraph{Scenario \#2: Repair Mode.}
Given an obsolete test case \(\tau = \left\langle V, S_{bk}, E, A\right\rangle \) containing broken stub code \(S_{bk}\), synthesize \(S\) to replace \(S_{bk}\) such that \(\testTuple \) passes.

In this scenario, our technique helps developers to repair test cases whose stub code is broken due to program or library changes.
For example, when the stub code in the highlighted part is broken because of code updates in \code{User}, \code{UserDao}, or \code{LoginService}, developers can specify the code lines that contain the broken stub code and \tool{} can replace the broken stub code with a synthesized one that can be compiled and can make the test pass.
Compared with scenario \#1, which synthesizes stub code from scratch, we leverage the information in the broken stub code \(S_{bk}\) to guide the synthesis in scenario \#2.
In this scenario, we target at the repair the stub code that are broken dur to refactoring or library upgrades.
It need developers to decide whether the stub code needs to be repaired.

\medskip{}
In both modes, \tool{} can be implemented as an IDE plugin.
In generation mode, developers can place the cursor at where the stub code needs to be generated.
In repair mode, developers can select the obsolete stub code that needs to be repaired.
Developers can place the cursor at where the stub code need to be generated or selecte the obsolete stub code that need to be repaired.
After that, they can launch \tool{} via a menu item provided by the plugin.
The stub code is synthesized to facilitate regression testing based on the current program version.
The test case with synthesized stub code captures the implemented behavior of the CUT, and helps to detect regression bugs in future versions of the CUT.
It is important to clarify that \tool{} does not guarantee the semantic equivalence between the synthesized stub code and developer-written stub code.
In the context of mocking, the stub code helping the test case achieving the same adequacy may not need to be syntactically or semantically similar.
As an example, Listing~\ref{lst:benchmark-36rep} shows the stub code written by developer and synthesized by \tool{}.
First, the developer-written stub code and the synthesized stub code are syntactically different.
The developer-written stub code is using the API \code{when(...).thenReturn(...)} while the synthesized stub code is using the API \code{doReturn(...).when(...)}.
Also, the semantics of the two stub codes are not exactly the same.
The developer-written stub will return \code{"/actuator/health"} only when the method \code{getPath} is invoked with an argument \code{"health"} while the synthesized stub code will return \code{"/actuator/health"} when the invocation is done with any argument.
Although they are neither syntactically nor semantically the same, the test cases with both stub codes execute exactly the same set of instructions, traverse exactly the same execution paths, and kill exactly the same set of mutants as shown in our evaluation (Table~5).
This is because in that test case, the method \code{getPath} will only be invoked with argument \code{"health"}.
In other words, the two stub codes are semantically equivalent in the context of that specific test case.

Indeed, \tool{} has available only the CUT executions \(E\) and assertions \(A\) to guide the generation and repair of stub code.
Such available information is unlikely to be a complete specification of the behavior of the test.
However, assertions should predicate of the salient expected behaviours of the test that makes the test pass or fail.
In our evaluation (Section ~\ref{sec:rq4}), we conjecture that obtaining stub code that fulfills such behaviors \ie{it makes the assertions pass} would be enough to achieve the same (or similar) test adequacy with the ground-truth stub code.

\subsection{Technical Challenges}\label{ssec:technical-challenges}

It is challenging to synthesize a stub code, say \(\mathcal{S}\), to pass an input test \(\tau\) due to the huge search space of possible candidates that can be generated for \(\tau\).
This is because a stub code candidate is free to stub any method of any mock object in \(M\) for an arbitrary number of times, and to return any value or throw any exception for each stub call.
Even by bounding the number of lines of code of the generated stub code (50 in our experiments), the search space is too huge to exhaustively explore.
However, only specific stub code \(\mathcal{S}\) can make \(\tau\) pass.
\(\mathcal{S}\) should stub the correct set of methods with proper values so that \(E\) executes without exceptions and all the constraints in \(A\) are satisfied.
Take the test case shown in Listing~\ref{lst:stubbing-example} as an example.
Lines~\ref{code:stubbing-example:act-begin}--\ref{code:stubbing-example:act-end} create a \code{LoginService} object with mock object \code{dao} and invoke \code{login} method with username ``foo'' and password ``bar''.
The test case passes only when satisfying two oracle assertions: (1) the \code{findUser} method is called twice (the mocking call at Line~\ref{code:stubbing-example:assert-begin}), and (2) the \code{loginResult} returned by \code{login} is successful (the assert statement at Line~\ref{code:stubbing-example:assert-end}).
Lines~\ref{code:stubbing-example:cut-begin}--\ref{code:stubbing-example:cut-end} show the implementation of the \code{login} method.
In this method, \code{dao.findUser} will be called twice only if it throws a \code{TimeoutException} when it is first called (executing Line~\ref{code:stubbing-example:catch}), and returns a \code{User} object when it is called for the second time (breaking the loop at Line~\ref{code:stubbing-example:loop}).
The \code{LoginResult} will be successful only if the \code{getPasswordHash} of the \code{User} object returns the SHA-1 checksum of password ``bar'' (passing the condition at Line~\ref{code:stubbing-example:hash-check}).
Lines~\ref{code:stubbing-example:stubbing-begin}--\ref{code:stubbing-example:arrange-end} show the specific stub code that makes the pass.

It is infeasible to identify the test-passing stub code \(\mathcal{S}\) by randomly or systematically exploring all the possible stub code candidates.
To address this problem, we propose to use an evolutionary algorithm to guide the synthesis of the stub code and search for \(\mathcal{S}\).

\paragraph{Key Idea.}
The evolutionary algorithm searches for the test-passing stub code \(\mathcal{S}\) by generating new candidate stub code via crossover and mutations of existing ones in a guided manner.
It guides the search by a fitness function that evaluates the distance between an arbitrary \(S\) and a passing stub code.
As discussed in Section~\ref{ssec:problem-statement}, \(\mathcal{S}\) is the set of stub code that can pass \(\tau\).
In other words, the stub code should (1) make the code in \(E\) executable, and (2) satisfy all oracle assertions in \(A\).
Based on this observation, we propose a multi-objective fitness function: with an arbitrary \(S\), we integrate it with \(\tau\) and capture
(1) the percentage of bytecode instructions in \(E\) that can be successfully executed, and (2) the percentage of oracle assertions in \(A\) that can be satisfied, and (3) the distance between the value outputted by \(S\) and its expected value for an unsatisfied oracle assertion.
Intuitively, \(S\) is closer to pass \(\tau\) if it can make more bytecode instructions in \(E\) executable, satisfy more oracle assertions in \(A\), and for the unsatisfied oracle assertions, the outputted value is closer to the expected value.

\section{\tool{}}\label{sec:approach}

\begin{figure}[t!]
	\centering
	\resizebox{0.75\linewidth}{!}{\includegraphics{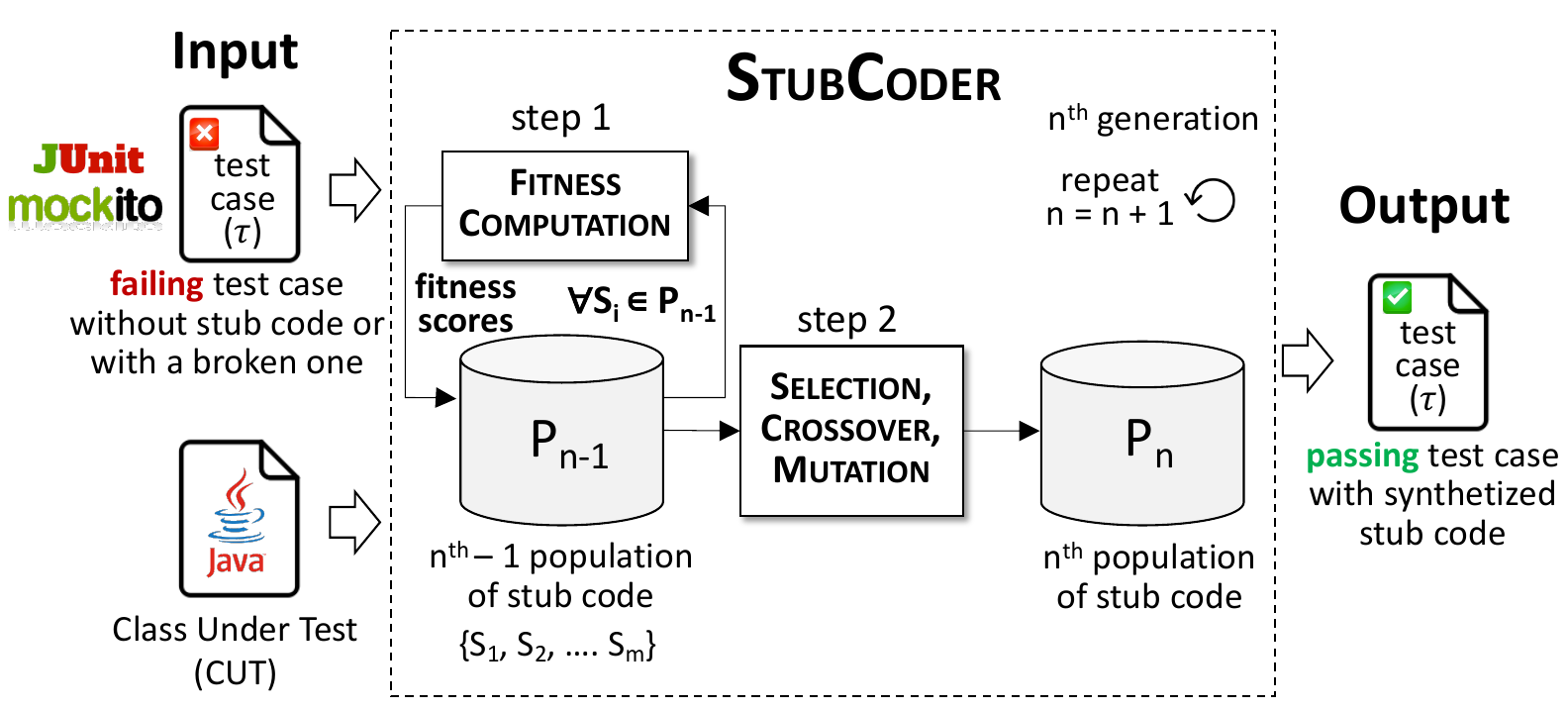}}
	\caption{Overview of \tool{}}\label{fig:overview}
\end{figure}

Figure~\ref{fig:overview} shows the logical architecture of \tool.
The input is a test case without stub code (\(\tau = \left\langle V,\varnothing, E, A\right\rangle \)) or with a broken one (\(\tau = \left\langle V,S_{bk}, E, A\right\rangle \)) and the corresponding CUT.
The output is the test case with a synthesized stub code that makes the test pass.
Specifically, \tool{} implements a population-based evolutionary algorithm that guides the search for stub code using a multi-objective fitness function, as discussed in Section~\ref{ssec:technical-challenges}.
At each generation, \tool{} evolves a population of stub code candidates until it finds one that can pass the test or the budget runs out.
Figure~\ref{fig:overview} shows the process of producing \(P_{i}\) the population at the \(i\)\textsuperscript{th} generation.
First, \tool{} computes the fitness score for each candidate individual (stub code) \(S \in P_{i-1}\). Then, it performs selection, crossover, and mutation to obtain the new population $P_i$.
In particular, the selection phase selects two parent individuals from \(P_{n-1}\). Individuals with higher fitness scores are more likely to be selected. The crossover phase combines the parents' genetic material (code elements in our case) to produce two offspring individuals. The mutation phase applies random mutations to the offspring individuals and adds them to $P_i$. These three phases repeat until $P_i$ is full.
In the following, we present the fitness function and explain how we adapt the selection, crossover, and mutation phases for the problem of stub code generation and repair.

\subsection{Fitness Function}
In this paper, we formulate the synthesis of stub code as a multi-objective optimization problem (MOOP)~\cite{DBLP:journals/tse/PanichellaOPL15,DBLP:conf/icec/TamakiKK96} with three objectives.
These objectives take into account the runtime behaviors of the Act and Assert phases of a test case.
Each of them focuses on a particular aspect of the runtime behavior of the test case \(\tau\) with the candidate stub code \(S\).

\paragraph{Stub Utilization (\(SU\)).}
A given stub code \(S\) can have multiple stub calls to specify the behaviors of the mock objects.
However, not necessarily all of the specified behaviors will be used during the Act phase.
For example, the CUT might not invoke a stubbed method or the argument does not match.
Such unused stub calls do not affect the behavior of the test, and mutating its return value has lower chances to make the test pass.
Therefore, we define stub effectiveness (SU) of a stub code \(S\) based on the number of stub calls that are used by the CUT during the Act phase, denoted by
\textit{used}$(S)$.
\[
	SU(S)=\tanh{\left(\frac{1}{C}{\textit{used}(S)} \right)}
\]
The hyperbolic tangent (\(\tanh\)) normalizes the value to \([0,1)\).
Since the curve of \(\tanh \) is more steep in the interval \([0,1)\) than in \([1, \infty)\), we divide the integer counter by a constant \(C > 1\) to make use of the range \([0,1)\).
In our experiment, we chose \(C=10\).

\paragraph{Exercise Coverage (\(EC\)).}
Apart from oracle assertion violations, a test case fails  when \(E\) invokes the CUT and the CUT throws an uncaught exception.
Such uncaught exceptions are caused by the incorrect behaviors specified by the stub code.
In general, a stub code that does not lead to uncaught exceptions when the test invokes the CUT is preferred, compared to one that does.
As such, we define the exercise coverage (EC) of a stub code \(S\) as the ratio of the executed bytecode instructions in \(E\).
\[
	EC(S)=\frac{|\left\{e \in E, e \text{ is executed}\right\}|}{|E|}
\]
\(EC\) penalizes the individuals with an early failure of \(E\) due to incorrect behaviors specified in the stub code.

\paragraph{Assertion Status (\(AS\)).}
AS is derived from the runtime behavior in the Assert phase where the test executes the assertions in \(A\).
It is computed from the score of each of the assertion oracles in \(A\) by taking their average.
\vspace{2mm}
\[
	AS(S)=\frac{1}{|A|}\cdot\sum_{a\in A}{score(a)}
\]
The score of each assertion oracle is a number in the range of \([0, 1]\), indicating how likely the assertion oracle is satisfied.
It is defined as the following.
\[
	score(a)=
	\begin{cases}
		1.0                                         & a \text{ is satisfied}                                     \\
		1.0 - d(a.\text{expected}, a.\text{actual}) & a \text{ is \code{assertEquals} and } a \text{ is failing} \\
		0.0                                         & \text{otherwise}                                           \\
	\end{cases}
\]
Specifically, for \junit{} \code{assertEquals} assertions, we measure the distance between the expected and actual values to estimate how far it is from being satisfied, which is similar to the branching condition distance~\cite{DBLP:journals/infsof/WegenerBS01}.
\[
	d(x, y)=
	\begin{cases}
		\tanh{\left(\frac{|x - y|}{|x|}\right)}   & x, y \text{ are numeric types}   \\
		\tanh{\left(\frac{Lev(x, y)}{|x|}\right)} & x, y \text{ are strings}         \\
		d(str(x), str(y))                         & x, y \text{ are complex objects} \\
	\end{cases}
\]
where \(Lev(x, y)\) is the Levenshtein distance~\cite{DBLP:journals/csur/Navarro01}.
Such a distance function considers the actions of insertions, deletions, and substitutions, which is in line with our mutation operators.
For complex objects, function \(str(x)\) serializes an object \(x\) into a string in a deep-copy manner~\cite{bloch2008effective}.
Specifically, it recursively converts the fields of complex objects into string representations.
This is because complex objects are often equated based on the values of its fields.
Therefore, we chose such an strategy to approximate the distance between two complex objects.
We implemented it using \code{ReflectionToStringBuilder} provided by the \textsc{Apache Commons} library~\cite{Tool:apachecommons}.
We fall back the denominator to 1.0 if it is zero to avoid division-by-zero error and further normalized the result into the range of \([0,1)\) with the hyperbolic tangent.

\begin{figure}[t]
\begin{lstlisting}[
		language=java,
		morekeywords={var},
		caption={A Test Case Adapted from Spring Boot Admin~\cite{GitHub:spb/cfaftest}},
		label={lst:benchmark-36},
		escapechar=|,
		linebackgroundcolor = {\ifnum \value{lstnumber} > 3 \ifnum \value{lstnumber} < 5 \color{stubbg} \fi \fi},
		numbers=left
	]
@Test
public void should_use_application_uri() {
    var endpoint = mock(PathMappedEndpoints.class);
    when(endpoint.getPath("health")).thenReturn("/actuator/health");
    var factory = new CFApplicationFactory(..., endpoint, ...);|\label{code:b36-act}|
    var app = factory.createApplication();|\label{code:b36-act-end}|
    assertEquals("base_url/actuator/health", app.getHealthUrl());|\label{code:b36-assert}|
}

// CUT
public Application createApplication() {
    return Application.builder().setHealthUrl(|\label{code:b36-cut}|
    "base_url" + endpoint.getPath("health")).build();|\label{code:b36-cut-end}|
}

//Candidate stub code
|\(S_a\)|: |\label{code:b36-sa}|
    when(endpoint.getPath("health")).thenReturn("random");|\label{code:b36-sa-end}|
|\(S_b\)|:|\label{code:b36-sb}|
    when(endpoint.getPath("health")).thenReturn("/actuator/hea");|\label{code:b36-sb-end}|
\end{lstlisting}
\end{figure}

\(AS\) can provide additional guidance to generate values that can satisfy oracle assertions specified with \code{assertEquals}.
For example, Listing~\ref{lst:benchmark-36} shows a code snippet adapted from a test in open-source project Spring Boot Admin~\cite{GitHub:spb/cfaftest}.
It tests the creation of an \code{Application} object from a factory class \code{CFApplicationFactory} (Lines~\ref{code:b36-act}--\ref{code:b36-act-end}).
The factory class sets up API end-points based on the information in an input \code{PathMappedEndPoints} that encapsulates a map from end-point names to their URLs.
In Lines \ref{code:b36-cut}--\ref{code:b36-cut-end}, the factory sets the application's \code{HealthUrl} by concatenating the base URL with the health URL encapsulated in \code{endpoint} by invoking its \code{getPath} method.
The oracle assertion of the test case verifies whether \code{HealthUrl} of the created application equals the string \code{"base\_url/actuator/health"} (Line~\ref{code:b36-assert}).
To make this test pass, the \code{getPath} method should be stubbed to return \code{"/actuator/health"} when it is invoked with argument \code{"health"}.
\(SU\), and \(EC\) cannot guide the generation of these specific values.
For example, candidate stub code \(S_a\) stubbing \code{getPath} to return \code{"random"} (Lines~\ref{code:b36-sa}--\ref{code:b36-sa-end}) and \(S_b\) stubbing \code{getPath} to return \code{"/actuator/hea"} (Lines~\ref{code:b36-sb}--\ref{code:b36-sb-end}) will achieve the same fitness score with only \(SU\) and \(EC\).
Both candidates can make the test execute to Line~\ref{code:b36-act-end} but violate the oracle assertion at Line~\ref{code:b36-assert}.
However, returning \code{"/actuator/hea"} is much closer to passing the test as the returned string is much more similar to the expected value \code{"/actuator/health"}.
This difference can be captured by \(AS\).

\paragraph{Fitness Computation.}
In this paper, we designed a dominance based fitness computation approach.
This is because, in our scenario, the three objectives are of different importance.
Based on the execution order of the arrange, act, and assert phases, there is a natural order of the three objectives: SU, EC, and AS.
Our rationale is that individuals who perform better (the functions have higher values) in later phases are more likely to converge to the test passing-stub code because in a test case, a later phase depends on the outcome of the former phases.
For example, the assert phase will be executed only when the arrange and act phases executed successfully.
Therefore, we define the dominance relationship \(\succ\) between as follows.
For two stub codes \(S_1\) and \(S_2\), \(S_1 \succ S_2\) if any of the following holds:
\begin{itemize}
	\item \(AS(S_1) > AS(S_2)\)
	\item \(AS(S_1) = AS(S_2) \wedge EC(S_1) > EC(S_2)\)
	\item \(AS(S_1) = AS(S_2) \wedge EC(S_1) = EC(S_2) \wedge SU(S_1) > SU(S_2)\)
\end{itemize}
During selection, for two individuals \ie{stub code} \(S_1\) and \(S_2\), we favor \(S_1\) if \(S_1 \succ S_2\).

\subsection{Representation of Stub Code}\label{sec:representation}

\begin{figure}[t]
	\centering
	\input{figures/stub_code_grammar.tex}
	\caption{Grammar of Synthesized Stub Code \(S\).}
	\label{fig:grammar}
\end{figure}

This section defines the possible stub code \(S\) that \tool{} is able to synthesize (the possible individuals of a population). It also describes how \tool{} represents an individual.
Specifically, a candidate stub code can be constructed by any of the possible strings on the context-free grammar shown in Figure~\ref{fig:grammar}.
Specifically, we represent \(S\) as a finite sequence of code elements, each of them can be either a \emph{variable definition} or a \emph{stub call}.
A variable definition constructs a value and stores it in a variable.
Then, a stub call can associate such variables with method calls on the mock objects.
These code elements in the stub code work together to specify the behavior of the mock objects.
By default, we set the length limit of \(S\) to 50, which is adequate for most of the stub code in practice (as shown in Table~\ref{tab:results}, the length of developer written stub code are less than 50).

\paragraph{Variable Definition.}
A variable definition \(v \leftarrow Expr\) defines a new variable \(v\) and initializes it with \(Expr\).
The \(Expr\) can be a literal value in \textsc{Java}~\cite{jls11}, an array of previously defined variables~\footnote{The synthesized stub code is inserted before the first reference of the mock objects in the test case. All the variables defined before the stub code can be used in the synthesized stub code.}, or an API call.
Specifically, an API call can be either a method call, a constructor call, or a field access, which also takes previously defined variables as arguments.
In addition, \(Expr\) can be the creation of a mock object.
This enables us to synthesize the mock objects that may be absent from the input~\eg{\code{user} in Listing~\ref{lst:stubbing-example}}.

\paragraph{Stub Call.}
A stub call specifies the reaction for a certain method call received by a mock object when the argument matcher matches all the arguments.
The reaction can be either \textsf{Return}\((v)\), which returns the value referenced by the variable \(v\), or \textsf{Throw}\((v)\), which throws the exception referenced by the variable \(v\).
There can be multiple stub calls matching the same method call on the same mock object.
Their reactions will be executed in the order that they appear.

\begin{figure}[t]
	\centering
	\footnotesize
	\begin{tikzpicture}[scale=0.5, node distance=0.2cm and 0cm]
    \tikzstyle{def} = [rectangle,draw]
    \tikzstyle{stub} = [rectangle,draw]
    \tikzstyle{code} = [rectangle]
    \tikzstyle{use} = [draw,->,-latex]

    \node[def] (Def0) {\(\mathbf{Def_1}\)};
    \node[code, right=of Def0] (Def0c) {\(v_0 \leftarrow \code{"foo"}\)};

    \node[def, below=of Def0] (Def1) {\(\mathbf{Def_1}\)};
    \node[code, right=of Def1] (Def1c) {\(v_1 \leftarrow \code{new TimeoutException()}\)};

    \node[stub, below=of Def1] (Stub1) {\(\mathbf{Stub_1}\)};
    \node[code, right=of Stub1] (Stub1c) {\(\langle\code{dao}, \code{findUser}, [\textsf{Eq}(v_0)] \rangle \to \textsf{Throw}(v_1)\)};

    \node[stub, below=of Stub1] (Stub2) {\(\mathbf{Stub_2}\)};
    \node[code, right=of Stub2] (Stub2c) {\(\langle\code{dao}, \code{findUser}, [\textsf{Eq}(v_0)]\rangle \to \textsf{Return}(\code{user})\)};

    \node[def, below=of Stub2] (Def2) {\(\mathbf{Def_2}\)};
    \node[code, right=of Def2] (Def3c) {\(v_2 \leftarrow \code{"bar"}\)};

    \node[def, below=of Def2] (Def3) {\(\mathbf{Def_3}\)};
    \node[code, right=of Def3] (Def4c) {\(v_3 \leftarrow \code{DigestUtils.sha1Hex(}v_2\code{)}\)};

    \node[stub, below=of Def3] (Stub3) {\(\mathbf{Stub_2}\)};
    \node[code, right=of Stub3] (Stub3c) {\(\langle \code{user}, \code{getPasswordHash}, \emptyset \rangle \to \textsf{Return}(v_3)\)};

    \path[use] (Def1) -- (Stub1);
    \path[use] (Def2) -- (Def3);
    \path[use] (Def3) -- (Stub3);
    \path[use] (Def0) [out=-140, in=140] to (Stub1);
    \path[use] (Def0) [out=-140, in=140] to (Stub2);
\end{tikzpicture}
	\caption{Representation of the Stub Code in Listing~\ref{lst:stubbing-example}}\label{fig:representation-example}
\end{figure}
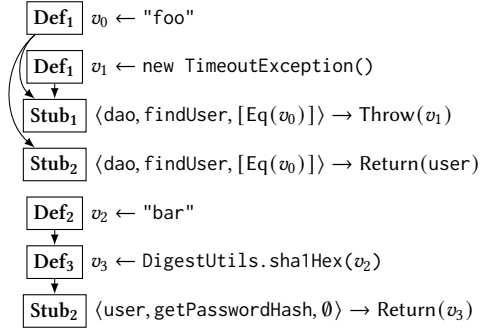

Figure~\ref{fig:representation-example} illustrates our representation of the stub code in Listing~\ref{lst:stubbing-example}.
There are four variable definitions and three stub calls.
As specified by \(Stub_1\) and \(Stub_2\), \code{dao.findUser} will throw a \code{TimeoutException} for the first call and return \code{user} for the second call.
The return value of \code{user.getPasswordHash} will be the SHA-1 digest of the string \code{"bar"} stored in \(v_3\).
The arrows indicate the def-use dependencies among these code elements.

The grammar in Figure~\ref{fig:grammar} is based on the APIs provided by the \mockito{}~\cite{Tool:mockito} framework, which is the most popular mocking framework for \java{}.
It can be adapted to support the syntax of other object-oriented programming languages and mocking frameworks.
Mocking frameworks tend to provide APIs with similar functionalities to aid the development.
For example, a \(StubCall\) can be mapped to a \code{expect(...)} call in \easymock{}~\cite{Tool:easymock} or a \code{Mock<T>.Setup(...)} call in \moq{}~\cite{Tool:moq4}.
\(\textsf{Return}(v)\) can be mapped to a \code{andReturn(...)} call in \easymock{} or a \code{MethodCall.Return(...)} call in \moq{}.
In our experiment, we implemented \tool{} in \java{} using \mockito{}.

\subsection{Evolutionary Algorithm}

Algorithm~\ref{alg:workflow} details the key steps in the evolution of stub code.
It takes a test case with a void or broken stub code \(\tau=\left\langle V,\varnothing, E,A\right\rangle\) or \(\left\langle V,S_{bk}, E,A\right\rangle \) as input, and outputs a stub code \(S\) such that the \(\testTuple \) is passing.
The evolution process is controlled by the population size \(N\) and generation budget \(MAX\_GEN\).

\paragraph{Symbol Pool.}
One challenge in synthesizing the stub code is to properly construct the return values.
However, it is less efficient to start searching from default values~\eg{0, \code{null}, an empty string} or random values.
To address this challenge, we construct a symbol pool to provide heuristics for the search process.
Specifically, the function \textsc{Construct-Symbol-Pool} in Algorithm~\ref{alg:workflow} extracts the literals and API calls from the \(\tau\) and the CUT.
It also includes the symbols in the broken stub code \(S_{bk}\), if available.
Such a constructed symbol pool contains useful values for synthesizing a test-passing stub code.
After construction, the symbol pool \(B\) is passed to \textsc{Crossover-and-Mutation}.
Mutation operators can take the literals and API calls in \(B\) to generate variable definitions.

\paragraph{Initial Population.}
\tool{} starts the synthesis process from an initial population.
Specifically, the function \textsc{Create-Initial-Population} returns a population of \(N\) stub code where each of them contains randomly generated code elements for the mock objects in \(M\).

\paragraph{Elitism Selection.}
Before starting populating the new population, \tool{} retains the best individuals.
At Line~\ref{alg:workflow:elitism} of Algorithm~\ref{alg:workflow}, the function \textsc{Elitism-Selection} selects the top 1\% of individuals with the highest fitness and brings them directly to the next generation.
With elitism selection, \tool{} avoids losing the best individuals in the next generation.

\paragraph{Parent Selection.}
At Line~\ref{alg:workflow:sel-parents} of Algorithm~\ref{alg:workflow}, \tool{} selects two parent individuals to produce the offspring.
In the function \textsc{Select-Parents}, we leverage tournament selection~\cite{DBLP:journals/compsys/MillerG95} to select the parents.
Tournament selection is widely used in genetic programming~\cite{DBLP:conf/issta/FraserZ10,DBLP:journals/tse/GouesNFW12} because it has less stochastic noise compared with other selection methods~\cite{DBLP:journals/ec/BlickleT96}.
Specifically, it randomly chooses \(K\) individuals and runs a tournament among them, after which the winner is chosen.
In this paper, we choose \(K=2\) to mitigate premature convergence and the local optimum problem~\cite{DBLP:conf/cec/LeggHK04,DBLP:conf/smc/LavinasASL18}.
Thus, we run two tournaments to get two parents \(\langle p_1, p_2\rangle\).

\begin{algorithm}[t]
	\SetFuncSty{textsc}
	\SetKwFunction{ConstructSymbolPool}{Construct-Symbol-Pool}
	\SetKwFunction{CreateInitialPop}{Create-Initial-Population}
	\SetKwFunction{Select}{Select-Parents}
	\SetKwFunction{CrossoverAndMutate}{Crossover-and-Mutate}
	\SetKwFunction{SelectBest}{Select-Best}
	\SetKwFunction{Elitism}{Elitism-Selection}
	\SetKwFunction{Fitness}{Fitness-Computation}
	\KwIn{Input test case \(\tau=\left\langle V,\varnothing, E,A\right\rangle \) or \(\left\langle V,S_{bk}, E,A\right\rangle \)}
	\KwIn{Population size \(N\), generation budget \(MAX\_GEN\)}
	\KwOut{Stub code \(S\)}

	\(B \leftarrow \) \ConstructSymbolPool{\(\tau\)} \;
	\(P \leftarrow \) \CreateInitialPop(\(N, B\)) \;
	\(P \leftarrow \Fitness{P}\)\;
	\(gen \leftarrow 1\) \;
	\Repeat{\(\exists S \in P, \tau=\left\langle V,S,E,A\right\rangle \) passes \(\vee gen > MAX\_GEN\)\label{alg:workflow:stop}}{
		\(P^\prime \leftarrow \Elitism{P} \) \label{alg:workflow:elitism}\;
		\While{\(|P^\prime| < N\)}{
			\(\langle p_1, p_2 \rangle \leftarrow \) \Select{\(P\)}\label{alg:workflow:sel-parents}\;
			\(\langle o_1, o_2  \rangle \leftarrow \) \CrossoverAndMutate{\(p_1, p_2, B\)}\;
			\(P^\prime \leftarrow P^\prime \cup \left\{ o_1, o_2 \right\} \) \;
		}
		\(P \leftarrow P^\prime\) \;
		\(P \leftarrow \Fitness{P}\)\;
		\(gen \leftarrow gen + 1\) \;

	}
	\Return{\(S\)}\;

	\caption{Evolution of Stub Code in \tool{}}\label{alg:workflow}
\end{algorithm}

\paragraph{Crossover and Mutation.}
The function \textsc{Crossover-and-Mutation} exchanges the genetic materials of two parents \(\langle p_1, p_2 \rangle \) and produces two offspring individuals \(\langle o_1, o_2 \rangle \), which are then mutated to introduce new genetic materials.

First, we exchange the stub calls that stub a mock object in \(M\) because they directly contribute to the outcome of the test case.
Specifically, we gather all such stub calls from \(p_1\) and \(p_2\), and copy each of them to \(o_1\) and \(o_2\) with probability 50\%.
In the stub code, a stub call relies on other code elements to function~\eg{the variable definition for its return value}.
Therefore, when copying a stub call, we perform a backward slicing~\cite{DBLP:journals/toplas/FerranteOW87} to obtain all its dependencies and copy all dependent code elements to the offspring.
For example, in Figure~\ref{fig:representation-example}, when copying \(Stub_2\), we bring together \(Def_3\) and \(Def_2\) since they are required to define \(v_3\), which is the return value of \(Stub_2\).

Next, the two offspring individuals \(o_1\) and \(o_2\) are mutated by one of the following mutation operators (randomly chosen with uniform probability).
\begin{itemize}
	\item \textit{\bfseries Inserting a Code Element.}
	      We randomly generate a stub call and a variable definition, and insert it into the stub code.
	      Variable definitions are generated by randomly choosing a literal or an API call from the symbol pool \(B\).
	\item \textit{\bfseries Altering the Parameters.}
	      Some code elements in the stub code take variables as their parameters.
	      For example, in Figure~\ref{fig:representation-example}, stub call \(Stub_1\) takes \(v_1\) to be the exception and variable definition \(Def_3\) uses \(v_2\) to perform an API call.
	      We randomly choose a parameter for such code elements and replace it with another variable of the same type in the stub code.
	\item \textit{\bfseries Altering the Literals.}
	      Some variable definitions in the stub code are numeric, string, or Boolean literals.
	      For example, \(Def_2\) in Figure~\ref{fig:representation-example} defines \(v_2\) with a string literal.
	      We apply a randomly chosen numeric or string operation \eg{add/subtract a random value, alter a character} to the literal.
	      For a Boolean literal, we simply flip it.
	\item \textit{\bfseries Swapping Two Code Elements.}
	      We randomly choose two code elements and interchange them.
	\item \textit{\bfseries Dropping a Code Element.}
	      We randomly remove a stub call or an unused variable definition from the stub code.
\end{itemize}

\paragraph{Mocking Decisions.}
The synthesized stub code can contain two types of mock objects.
The mocking decision is made depending on how they are declared.
\begin{itemize}
	\item \textit{\bfseries Type I Mock Objects.}
	      For the mock objects declared by developers in the test cases, we follow the original mocking decision by the developers.
	\item \textit{\bfseries Type II Mock Objects.}
	      During mutation, using a real object or a mock object is an alternative way of generating an object, and StubCoder will choose randomly between these alternative ways.
	      For instance, when populating a variable of a complex type \(T\), the mutation operator will randomly choose between using a generator of \(T\) \eg{the constructors of \(T\), the fields of type \(T\), and the methods that return \(T\)}, and a mocked version of \(T\) \ie{\code{mock(T.class)}}.
	      Since the goal of \tool{} is to synthesize a stub code to pass the test, the fitness function will then prioritize the stub code that is more likely to make the test pass.
\end{itemize}

\paragraph{Stopping Criterion.}
As shown at Line~\ref{alg:workflow:stop} of Algorithm~\ref{alg:workflow}, we stop the algorithm if the best stub code makes the test pass or the maximum number of generations is reached.

\section{Evaluation}\label{sec:evaluation}
This section presents the evaluation of~\tool{}.
Specifically, we aim to answer the following four research questions.
\begin{itemize}
	\item \textbf{RQ1 (Stub Code Generation):}
	      \textit{Is \tool{} effective in generating stub code?}
	\item \textbf{RQ2 (Stub Code Repair):}
	      \textit{
		      Is \tool{} effective in repairing obsolete test cases due to broken stub code?
		      Does it outperform state-of-the-art program repair techniques?
	      }
	\item \textbf{RQ3 (Effectiveness of the Fitness Function):} \textit{Can the fitness function effectively guide stub code synthesis?}
	\item \textbf{RQ4 (Fidelity of the Synthesized Stub Code):} \textit{To what extend does the stub code synthesized by \tool{} preserve the effect of the ground-truth stub code?}
\end{itemize}
\smallskip
RQ1 and RQ2 evaluate \tool{} in the two application scenarios (synthesis and repair of stub code, see Section~\ref{ssec:problem-statement}).
RQ3 evaluates the effectiveness of \tool's fitness function by comparing it with an unguided (random) strategy.
RQ4 compares the fidelity of the stub code synthesized by \tool{} with that written by developers.
Specifically, we evaluate to what extent the test case with synthesized stub code can preserve the runtime effect of the test case with the ground-truth stub code by comparing their executed instructions, execution paths, and ability to kill mutants.

\subsection{Evaluation Subjects}

\paragraph{Benchmark Description.}
To answer the four research questions, we constructed a benchmark of \numOfBenchmarkEntries{} real-world test cases selected from \numOfBenchmarkProjects{} open-source projects (see Table~\ref{tab:subjects}). Each entry in the benchmark contains:
\begin{itemize}
	\item The test case with the removed stub code, which is the input of \tool{} in RQ1.
	\item The broken version of the stub code, which is the additional input of \tool{} in RQ2.
	\item The ground-truth stub code written by developers that makes the test pass, which is used for comparison in RQ4.
	\item The production code and the dependent libraries, which are required to compile and run the test.
\end{itemize}

\paragraph{Project Selection.}
To build the benchmark, we searched on GitHub~\cite{github} for open-source \textsc{Java} projects and sorted the results by the number of stars, which is an indicator of popularity.
We manually went through the top 150 projects to identify those meeting the four criteria below:

\begin{enumerate}
	\item It has at least 1,000 lines of \textsc{Java} code.
	      This is to filter out small projects.
	\item It uses the \mockito{} framework~\cite{Tool:mockito} to simulate/verify the behaviors of test dependencies.
	      This is because we implemented \tool{} based on \mockito{}, which is the most popular mocking framework for \textsc{Java}~\cite{DBLP:conf/qsic/MostafaW14}.
	\item It is not an \textsc{Android} project since \textsc{Android} is currently not supported by our implementation.
	\item It uses \textsc{Maven} or \textsc{Gradle} as build automation tools so that we can automate the dependency collection procedure.
\end{enumerate}
We identified 40 candidate projects that satisfy our selection criteria.

\paragraph{Benchmark Preparation.}
For each candidate project, we automatically explored their commit history since 2018.
We performed an AST-level diff using \textsc{GumTree}~\cite{DBLP:conf/kbse/FalleriMBMM14} between the two versions of each commit to locate changes to the \mockito stub code.
The diff returned 2,295 code changes.
Since preparing the benchmark requires intensive manual effort, we performed a pre-selection on the code changes.
For each of the projects, we sampled at most 100 code changes, obtaining a total of 871 candidates.
Then, we manually read the code diff and commit messages to understand the semantics of the change.
We ignored the code changes that simply rename code elements~\ie{\textsc{GumTree} classifies the ASTs before and after the changes as isomorphic}.
This is because we do not regard such trivial code changes as the target application scenario of \tool{}.
Repairing stub code in such cases can be easily achieved using the refactoring feature of modern IDEs.

\begin{figure}[t]
	\centering
	\resizebox{0.65\linewidth}{!}{\includegraphics{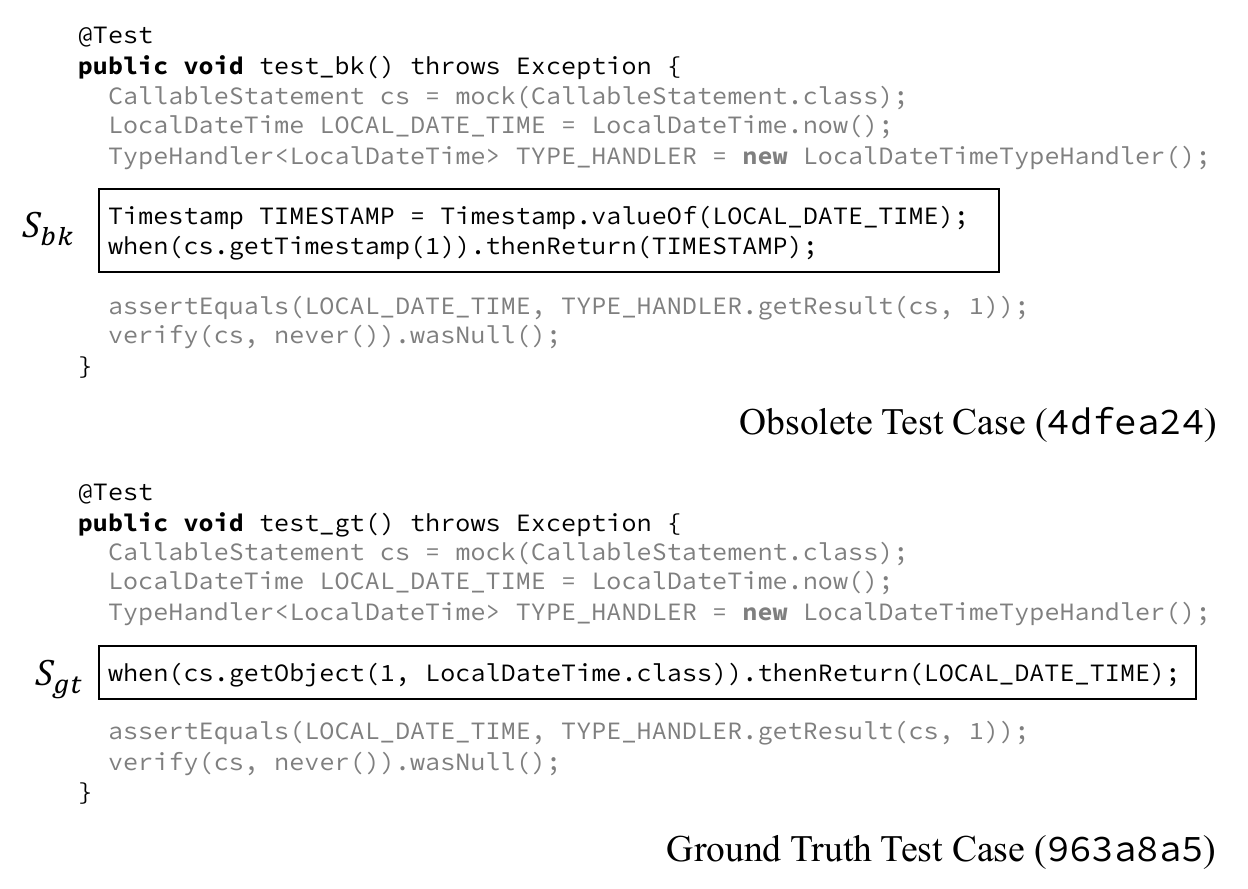}}
	\caption{Version Relationship between \(S_{bk}\) and \(S_{gt}\) in Project MB3\label{fig:version-diagram}}
\end{figure}

After dropping the trivial cases and duplicate commits due to git branch merges, we retained 261 code changes.
Each code change specifies two versions of a stub code: a broken one (before the change), and a correct one (after the change).
To turn each code change into a benchmark entry, we performed the following procedure:

\begin{itemize}
	\item We ran the \textsc{Gradle} or \textsc{Maven} build script to resolve the dependencies and compile the project.
    \item We rewrote each oracle assertion written with custom assertion frameworks into semantically equivalent \junit{} assertions.
	      Table~\ref{tab:rewritten-assertions} lists the rewritten rules applied by us.
	      Specifically, the rules were drafted by one author and then independently validated by two other authors independently.
	      One more author joined and resolved disagreements when they occurred.
	      In addition, we ran the test cases after rewriting the assertions to check that the test is still passing.
	      This was done for 14 subjects in project AZK, SBA, JIB, GRC, ZKN, and SPB.
	\item We executed the test case to ensure that it passes with the correct stub code written by developers, since we will use it as the ground truth.
	\item We removed the stub code so that the test fails. This is to ensure that the stub code is required to pass the test.
\end{itemize}
We discarded a code change if we failed to perform any of the steps above on it.
Finally, we constructed a benchmark of \numOfBenchmarkEntries{} test cases containing 167 mock objects collected from \numOfBenchmarkProjects{} projects.
Each of the entries in the benchmark consists of two elements \(\left\langle \tau_{bk}, \tau_{gt}\right\rangle\).
\(\tau_{bk} = \langle V, S_{bk}, E, A\rangle\) is the obsolete test case containing broken stub code \(S_{bk}\) and \(\tau_{gt} = \langle V, S_{gt}, E, A\rangle\) is the fixed version of the test case, which contains the ground-truth stub code \(S_{gt}\) written by developers.
Figure~\ref{fig:version-diagram} shows an example of a benchmark entry in project MB3.
The obsolete test case from version \code{4dfea24} contains broken stub code.
Developers fixed the broken stub code in version \code{963a8a5} by modifying the stub code.
Table~\ref{tab:subjects} summarizes the benchmark.

\begin{table}[]
	\caption{List of Rewritten Assertions in Benchmark}\label{tab:rewritten-assertions}
	\centering
	\footnotesize
	\renewcommand{\arraystretch}{0.9}
	\begin{tabular}{ll}
    \toprule
    \textbf{Original Assertion} & \textbf{Rewritten \junit{} Version} \\
    \midrule
    \code{assertThat(x, equalTo(y))} & \code{assertEquals(y, x)} \\ 
    \code{assertThat(x).isEqualTo(y)} & \code{assertEquals(y, x)} \\ 
    \code{assertThat(x).isNotNull()} & \code{assertNotNull(x)} \\ 
    \code{assertThat(x).isInstanceOf(Y.class)} & \code{assertTrue(x instanceof Y)} \\ 
    \code{assertThat(x).isSameAs(y)} & \code{assertSame(y, x)} \\ 
    \midrule
    \begin{tabular}{@{}l@{}}
        \code{assertThatThrownBy(() -> x)}\\
        \code{\ \ .isInstanceOf(E.class).hasMessage(y)}
    \end{tabular}
    &
    \begin{tabular}{@{}l@{}}
        \code{E e = assertThrows(E.class, () -> x)}\\
        \code{assertEquals(y, e.getMessage())}
    \end{tabular}
    \\
    \midrule
    \begin{tabular}{@{}l@{}}
        \code{assertThatThrownBy(() -> x)}\\
        \code{\ \ .isInstanceOf(E.class).hasMessageStartsWith(y)}
    \end{tabular}
    &
    \begin{tabular}{@{}l@{}}
        \code{E e = assertThrows(E.class, () -> x)}\\
        \code{assertTrue(e.getMessage().startsWith(y))}
    \end{tabular}
    \\
    \bottomrule
\end{tabular}
\end{table}

\begin{table}[]
	\caption{Demographics of Benchmark}\label{tab:subjects}
	\centering
	\footnotesize
	\renewcommand{\arraystretch}{0.9}
	\begin{tabularx}{\linewidth}{lXrrrr}
    \toprule
    \textbf{Artifact ID} & \textbf{GitHub Project ID}            & \textbf{LOC (Java)}    &  \textbf{Stars}  & \textbf{\# Test Cases} & \textbf{\# Total Mock Objects} \\
    \midrule
    ADR                  & \code{apache/druid}                   & 856K             &  11.9K           & 2                      & 7                              \\
    ADU                  & \code{apache/dubbo}                   & 199K             &  37.6K           & 13                     & 25                             \\
    AHP                  & \code{apache/hadoop}                  & 1,834K           &  12.7K           & 5                      & 9                              \\
    AZK                  & \code{apache/zookeeper}               & 116K             &  10.6K           & 1                      & 2                              \\
    APL                  & \code{apolloconfig/apollo}            & 52K              &  27.0K           & 8                      & 40                             \\
    SBA                  & \code{codecentric/spring-boot-admin}  & 18K              &  11.0K           & 2                      & 2                              \\
    JIB                  & \code{GoogleContainerTools/jib}       & 55K              &  11.9K           & 2                      & 7                              \\
    GRC                  & \code{grpc/grpc-java}                 & 226K             &  9.9K            & 3                      & 3                              \\
    MB3                  & \code{mybatis/mybatis-3}              & 68K              &  17.4K           & 12                     & 37                             \\
    N4J                  & \code{neo4j/neo4j}                    & 731K             &  10.2K           & 4                      & 25                             \\
    ZUL                  & \code{Netflix/zuul}                   & 232K             &  12.0K           & 1                      & 2                              \\
    ZKN                  & \code{openzipkin/zipkin}              & 42K              &  15.5K           & 5                      & 7                              \\
    SPB                  & \code{spring-projects/spring-boot}    & 347K             &  62.1K           & 1                      & 1                              \\
    \midrule                  
    \multicolumn{2}{l}{\textbf{Total}}                           &                  &                  & 59                     & 167                            \\
    \bottomrule
\end{tabularx}
\end{table}

\begin{table}[]
	\centering
	\caption{RQ1 --- RQ3: Comparison of the Success Rate in Different Setups}\label{tab:results}
	\vspace{-1.2em}
	\captionsetup{style=base,singlelinecheck=off,font=small}
	\caption*{\scriptsize
		GT is for ground truth, which are the test cases written by developers.
		``\tool{} (G)'',  ``\tool{} (R)'', ``NSGA-II'', ``Weighted Sum'', and ``Unguided'' are \tool{} in generation mode, repair mode, with NSGA-II, with weighted sum, and random selection, respectively.
		\(|M|\) is the number of mock objects in \(V\) in the test case.
        \(|A|\) is the number of assertions (including \code{verify} assertions on mock objects) in the test case.
		\(|S|\) is the size of the stub code.
		SR is the number of successful runs.
		Gen, Time, and \(|S|\) are the number of generations taken, time taken (in seconds), and the size of stub code, respectively.
	}
	\vspace{-1em}
	\scriptsize
	\renewcommand{\arraystretch}{0.9}
	\begin{center}
    \setlength{\tabcolsep}{2.3pt}
    \begin{tabularx}{\linewidth}{ll|rrr|rrrr|rrrr|rrrr|rrrr|rrrr}
        \toprule
        \multicolumn{2}{l}{\multirow{3}{*}{\textbf{Subject}}}
        & \multicolumn{3}{c}{\textbf{GT}}
        & \multicolumn{4}{c}{\textbf{\tool{} (G)}}
        & \multicolumn{4}{c}{\textbf{\tool{} (R)}}
        & \multicolumn{4}{c}{\textbf{NSGA-II}}
        & \multicolumn{4}{c}{\textbf{Weighted Sum}}
        & \multicolumn{4}{c}{\textbf{Unguided}} \\
        \cmidrule(l{1pt}r{1pt}){3-5}
        \cmidrule(l{1pt}r{1pt}){6-9}
        \cmidrule(l{1pt}r{1pt}){10-13}
        \cmidrule(l{1pt}r{1pt}){14-17}
        \cmidrule(l{1pt}r{1pt}){18-21}
        \cmidrule(l{1pt}r{1pt}){22-25}
        && \(|M|\) & \(|A|\) & \(|S|\)&
        SR & Gen & Time & \(|S|\) &
        SR & Gen & Time & \(|S|\) &
        SR & Gen & Time & \(|S|\) &
        SR & Gen & Time & \(|S|\) &
        SR & Gen & Time & \(|S|\) \\
        \midrule[\heavyrulewidth]
        \multirow{4}{*}{N4J}  & \#1  & 1  & 1  & 3   &       10 & 9   & 511   & 3  &      10 & 13  & 749   & 3  &       4  & 335 & 26155 & 3  &      10 & 8   & 534   & 3  &        10 & 25  & 983   & 3    \\
                              & \#2  & 5  & 4  & 26  &       10 & 24  & 1118  & 11 &      10 & 10  & 454   & 20 &       10 & 27  & 1670  & 4  &      10 & 14  & 861   & 9  &        10 & 37  & 1892  & 4    \\
                              & \#3  & 9  & 5  & 11  &       0  & -   & -     & -  &      2  & 342 & 16738 & 9  &       0  & -   & -     & -  &      1  & 387 & 23048 & 18 &        0  & -   & -     & -    \\
                              & \#4  & 10 & 3  & 9   &       9  & 75  & 4949  & 17 &      8  & 92  & 5882  & 20 &       0  & -   & -     & -  &      10 & 149 & 11294 & 39 &        1  & 307 & 20908 & 12   \\
\midrule                SPB   & \#5  & 1  & 1  & 1   &       10 & 1   & 38    & 2  &      10 & 1   & 36    & 2  &       10 & 1   & 45    & 2  &      10 & 1   & 56    & 2  &        10 & 1   & 40    & 2    \\
\midrule\multirow{3}{*}{GRC}  & \#6  & 1  & 2  & 1   &       10 & 2   & 69    & 2  &      10 & 2   & 67    & 2  &       10 & 2   & 95    & 2  &      10 & 2   & 94    & 2  &        2  & 156 & 6254  & 2    \\
                              & \#7  & 1  & 2  & 14  &       0  & -   & -     & -  &      0  & -   & -     & -  &       0  & -   & -     & -  &      0  & -   & -     & -  &        0  & -   & -     & -    \\
                              & \#8  & 1  & 2  & 12  &       0  & -   & -     & -  &      0  & -   & -     & -  &       0  & -   & -     & -  &      0  & -   & -     & -  &        0  & -   & -     & -    \\
\midrule\multirow{12}{*}{MB3} & \#9  & 4  & 2  & 1   &       10 & 2   & 86    & 1  &      10 & 1   & 43    & 2  &       10 & 2   & 97    & 1  &      10 & 1   & 51    & 1  &        10 & 1   & 43    & 1    \\
                              & \#10 & 2  & 2  & 1   &       10 & 2   & 82    & 1  &      10 & 2   & 85    & 2  &       10 & 1   & 54    & 1  &      10 & 2   & 96    & 1  &        10 & 1   & 40    & 1    \\
                              & \#11 & 4  & 2  & 1   &       10 & 2   & 89    & 1  &      10 & 3   & 131   & 1  &       10 & 3   & 145   & 1  &      10 & 3   & 196   & 1  &        10 & 2   & 91    & 2    \\
                              & \#12 & 4  & 2  & 1   &       10 & 2   & 81    & 3  &      10 & 2   & 85    & 1  &       10 & 2   & 95    & 2  &      10 & 2   & 136   & 4  &        10 & 1   & 46    & 1    \\
                              & \#13 & 1  & 2  & 1   &       10 & 2   & 76    & 1  &      10 & 2   & 77    & 1  &       10 & 1   & 43    & 2  &      10 & 1   & 57    & 1  &        10 & 1   & 40    & 1    \\
                              & \#14 & 4  & 2  & 1   &       10 & 2   & 89    & 1  &      10 & 1   & 42    & 1  &       10 & 5   & 237   & 1  &      10 & 1   & 68    & 1  &        10 & 2   & 90    & 2    \\
                              & \#15 & 4  & 2  & 1   &       10 & 3   & 127   & 1  &      10 & 2   & 86    & 2  &       10 & 2   & 98    & 1  &      10 & 1   & 55    & 2  &        10 & 1   & 44    & 1    \\
                              & \#16 & 4  & 2  & 1   &       10 & 2   & 85    & 1  &      10 & 3   & 128   & 1  &       10 & 4   & 203   & 1  &      10 & 4   & 218   & 1  &        10 & 1   & 44    & 1    \\
                              & \#17 & 1  & 2  & 1   &       10 & 2   & 78    & 1  &      10 & 4   & 158   & 1  &       10 & 2   & 91    & 2  &      10 & 2   & 114   & 1  &        10 & 1   & 42    & 2    \\
                              & \#18 & 4  & 2  & 1   &       10 & 2   & 85    & 1  &      10 & 5   & 219   & 2  &       10 & 2   & 99    & 1  &      10 & 3   & 159   & 1  &        10 & 1   & 45    & 1    \\
                              & \#19 & 1  & 2  & 1   &       10 & 2   & 79    & 1  &      10 & 4   & 156   & 1  &       10 & 4   & 218   & 1  &      10 & 1   & 47    & 2  &        10 & 1   & 41    & 1    \\
                              & \#20 & 4  & 2  & 1   &       10 & 5   & 210   & 1  &      10 & 2   & 86    & 1  &       10 & 3   & 145   & 1  &      10 & 2   & 111   & 1  &        10 & 1   & 46    & 1    \\
\midrule\multirow{5}{*}{ZKN}  & \#21 & 1  & 2  & 7   &       10 & 1   & 35    & 2  &      10 & 1   & 37    & 2  &       10 & 13  & 567   & 2  &      10 & 1   & 56    & 2  &        10 & 1   & 38    & 2    \\
                              & \#22 & 2  & 1  & 8   &       0  & -   & -     & -  &      0  & -   & -     & -  &       0  & -   & -     & -  &      0  & -   & -     & -  &        0  & -   & -     & -    \\
                              & \#23 & 1  & 2  & 7   &       10 & 25  & 953   & 3  &      10 & 20  & 745   & 3  &       0  & -   & -     & -  &      10 & 14  & 313   & 5  &        0  & -   & -     & -    \\
                              & \#24 & 2  & 1  & 7   &       0  & -   & -     & -  &      0  & -   & -     & -  &       0  & -   & -     & -  &      0  & -   & -     & -  &        0  & -   & -     & -    \\
                              & \#25 & 1  & 2  & 10  &       10 & 10  & 381   & 6  &      10 & 18  & 689   & 4  &       0  & -   & -     & -  &      10 & 20  & 428   & 3  &        0  & -   & -     & -    \\
\midrule\multirow{8}{*}{APL}  & \#26 & 5  & 3  & 3   &       10 & 2   & 78    & 2  &      10 & 3   & 121   & 2  &       10 & 3   & 133   & 2  &      10 & 2   & 131   & 4  &        9  & 61  & 2646  & 3    \\
                              & \#27 & 5  & 4  & 3   &       10 & 2   & 79    & 5  &      10 & 2   & 80    & 6  &       10 & 2   & 90    & 4  &      10 & 2   & 103   & 3  &        10 & 2   & 86    & 3    \\
                              & \#28 & 6  & 2  & 10  &       10 & 111 & 5116  & 9  &      9  & 85  & 3895  & 9  &       7  & 158 & 8409  & 10 &      9  & 84  & 6597  & 11 &        1  & 324 & 16501 & 10   \\
                              & \#29 & 4  & 2  & 10  &       1  & 65  & 2777  & 12 &      4  & 311 & 13022 & 18 &       1  & 143 & 7282  & 17 &      0  & -   & -     & -  &        0  & -   & -     & -    \\
                              & \#30 & 6  & 5  & 20  &       5  & 80  & 3673  & 11 &      5  & 233 & 10805 & 6  &       5  & 151 & 7906  & 7  &      7  & 279 & 14574 & 19 &        0  & -   & -     & -    \\
                              & \#31 & 6  & 11 & 24  &       0  & -   & -     & -  &      0  & -   & -     & -  &       0  & -   & -     & -  &      0  & -   & -     & -  &        0  & -   & -     & -    \\
                              & \#32 & 4  & 4  & 10  &       10 & 1   & 41    & 2  &      10 & 1   & 39    & 2  &       10 & 1   & 45    & 10 &      10 & 1   & 53    & 6  &        10 & 1   & 52    & 8    \\
                              & \#33 & 4  & 3  & 12  &       10 & 26  & 1196  & 3  &      10 & 22  & 949   & 5  &       9  & 98  & 4980  & 4  &      10 & 17  & 918   & 5  &        0  & -   & -     & -    \\
\midrule                ZUL   & \#34 & 2  & 1  & 2   &       10 & 1   & 39    & 1  &      10 & 1   & 37    & 1  &       10 & 6   & 273   & 1  &      10 & 1   & 51    & 1  &        10 & 1   & 51    & 3    \\
\midrule\multirow{2}{*}{SBA}  & \#35 & 1  & 3  & 1   &       0  & -   & -     & -  &      10 & 17  & 639   & 2  &       1  & 72  & 3095  & 2  &      0  & -   & -     & -  &        0  & -   & -     & -    \\
                              & \#36 & 1  & 3  & 1   &       0  & -   & -     & -  &      10 & 5   & 186   & 2  &       0  & -   & -     & -  &      0  & -   & -     & -  &        0  & -   & -     & -    \\
\midrule\multirow{2}{*}{JIB}  & \#37 & 5  & 18 & 13  &       0  & -   & -     & -  &      0  & -   & -     & -  &       0  & -   & -     & -  &      0  & -   & -     & -  &        0  & -   & -     & -    \\
                              & \#38 & 2  & 1  & 15  &       0  & -   & -     & -  &      2  & 379 & 15510 & 32 &       0  & -   & -     & -  &      0  & -   & -     & -  &        0  & -   & -     & -    \\
\midrule                AZK   & \#39 & 2  & 20 & 11  &       0  & -   & -     & -  &      0  & -   & -     & -  &       0  & -   & -     & -  &      0  & -   & -     & -  &        0  & -   & -     & -    \\
\midrule\multirow{5}{*}{AHP}  & \#40 & 1  & 1  & 9   &       10 & 1   & 43    & 4  &      10 & 1   & 43    & 2  &       10 & 1   & 50    & 2  &      10 & 1   & 62    & 8  &        10 & 1   & 44    & 2    \\
                              & \#41 & 1  & 3  & 9   &       10 & 7   & 312   & 6  &      10 & 8   & 364   & 7  &       10 & 5   & 237   & 10 &      10 & 7   & 380   & 7  &        10 & 3   & 141   & 10   \\
                              & \#42 & 3  & 3  & 20  &       10 & 1   & 66    & 6  &      10 & 1   & 66    & 6  &       10 & 2   & 143   & 6  &      10 & 1   & 87    & 13 &        10 & 1   & 66    & 5    \\
                              & \#43 & 2  & 3  & 7   &       8  & 27  & 1252  & 14 &      10 & 17  & 780   & 34 &       7  & 100 & 4827  & 14 &      10 & 30  & 1695  & 21 &        10 & 77  & 3577  & 13   \\
                              & \#44 & 2  & 4  & 8   &       10 & 32  & 1487  & 13 &      10 & 75  & 3553  & 12 &       4  & 203 & 9984  & 9  &      10 & 29  & 1648  & 13 &        10 & 56  & 2831  & 13   \\
\midrule\multirow{2}{*}{ADR}  & \#45 & 1  & 1  & 1   &       2  & 232 & 12684 & 2  &      0  & -   & -     & -  &       0  & -   & -     & -  &      1  & 169 & 10534 & 2  &        0  & -   & -     & -    \\
                              & \#46 & 6  & 6  & 16  &       10 & 5   & 482   & 9  &      10 & 5   & 391   & 8  &       10 & 4   & 257   & 4  &      10 & 4   & 506   & 10 &        10 & 4   & 212   & 5    \\
\midrule\multirow{13}{*}{ADU} & \#47 & 2  & 1  & 3   &       7  & 109 & 4404  & 17 &      2  & 174 & 7052  & 15 &       0  & -   & -     & -  &      8  & 112 & 5413  & 18 &        0  & -   & -     & -    \\
                              & \#48 & 3  & 3  & 10  &       9  & 41  & 1805  & 9  &      8  & 71  & 3234  & 10 &       9  & 120 & 5712  & 21 &      9  & 91  & 4713  & 20 &        0  & -   & -     & -    \\
                              & \#49 & 1  & 2  & 4   &       10 & 54  & 2118  & 7  &      10 & 25  & 966   & 9  &       0  & -   & -     & -  &      10 & 72  & 3411  & 6  &        0  & -   & -     & -    \\
                              & \#50 & 1  & 2  & 4   &       10 & 80  & 3153  & 8  &      10 & 15  & 552   & 10 &       0  & -   & -     & -  &      10 & 62  & 3320  & 8  &        0  & -   & -     & -    \\
                              & \#51 & 2  & 1  & 7   &       6  & 46  & 1878  & 12 &      3  & 73  & 3053  & 19 &       6  & 303 & 14601 & 8  &      9  & 36  & 1844  & 7  &        0  & -   & -     & -    \\
                              & \#52 & 1  & 1  & 2   &       10 & 1   & 35    & 1  &      10 & 1   & 35    & 1  &       10 & 1   & 38    & 1  &      10 & 1   & 59    & 1  &        10 & 1   & 37    & 1    \\
                              & \#53 & 1  & 2  & 9   &       10 & 77  & 3191  & 14 &      10 & 50  & 2143  & 12 &       1  & 384 & 15915 & 15 &      10 & 88  & 4661  & 29 &        0  & -   & -     & -    \\
                              & \#54 & 8  & 2  & 14  &       8  & 61  & 2733  & 28 &      8  & 183 & 8284  & 33 &       0  & -   & -     & -  &      5  & 135 & 7114  & 23 &        0  & -   & -     & -    \\
                              & \#55 & 1  & 1  & 2   &       10 & 1   & 36    & 1  &      10 & 1   & 35    & 1  &       10 & 1   & 40    & 1  &      10 & 1   & 60    & 1  &        10 & 1   & 38    & 1    \\
                              & \#56 & 1  & 1  & 5   &       10 & 43  & 1848  & 9  &      10 & 32  & 1291  & 14 &       1  & 378 & 16034 & 9  &      10 & 29  & 1512  & 10 &        10 & 196 & 7896  & 15   \\
                              & \#57 & 1  & 2  & 7   &       1  & 223 & 11755 & 7  &      3  & 373 & 12003 & 6  &       0  & -   & -     & -  &      3  & 136 & 7933  & 6  &        0  & -   & -     & -    \\
                              & \#58 & 1  & 1  & 3   &       9  & 25  & 878   & 7  &      10 & 27  & 1047  & 7  &       5  & 215 & 8896  & 7  &      10 & 26  & 1400  & 6  &        9  & 52  & 2416  & 12   \\
                              & \#59 & 2  & 1  & 7   &       10 & 2   & 73    & 8  &      10 & 2   & 72    & 5  &       5  & 79  & 2691  & 2  &      10 & 3   & 171   & 8  &        10 & 4   & 175   & 16   \\
        \midrule[\heavyrulewidth]
        \multicolumn{5}{l}{\textbf{Success Rate > 50\%}}&  \multicolumn{4}{|l}{76\%}  &  \multicolumn{4}{|l}{76\%} &  \multicolumn{4}{|l}{58\%}  &  \multicolumn{4}{|l}{76\%} &    \multicolumn{4}{|l}{54\%}   \\
        \bottomrule
    \end{tabularx}
\end{center}

\end{table}

\subsection{RQ1: Stub Code Generation}\label{sec:rq1}

\paragraph{Experiment Setup.}
To answer RQ1, we ran \tool{} on the \numOfBenchmarkEntries{} test cases without stub code in our benchmark to generate stub code for them.
We ran \tool{} with population size \(N=200\), and set the generation budget \(MAX\_GEN = 400\).
We selected these parameters based on a few trial runs following previous work~\cite{DBLP:conf/sigsoft/TerragniJTP20}.
Due to the stochastic nature of evolutionary algorithms, we evaluated whether \tool{} can successfully synthesize the stub code in 10 repetitions.
Our algorithm relies on a pseudo-random number generator when making random decisions.
We chose 10 randomly-generated prime numbers as the random seeds for each repetitions, these random seeds are used across all the evaluation subjects.
We also designed two alternative optimization strategies in addition to our dominance based approach for comparison.
\begin{itemize}
	\item \textbf{Weighted Sum.}
	      The first alternative optimization strategy is to combine these objectives using weighted sum.
	      In this setup, we combined the three objectives into a single fitness function with different weights.
	      With the same rationale as for the NSGA-II variant, we assigned higher weights for the functions measuring the later stages of test execution, following the powers of 2.
	      \[
		      \text{\textbf{fitness}}(S) = 2^0\cdot SU(S)+ 2^1\cdot EC(S) + 2^2 \cdot AS(S)
	      \]
	\item \textbf{NSGA-II.}
	      In addition, we tried out the most popularly adopted MOEA, NSGA-II~\cite{DBLP:journals/tec/DebAPM02}.
	      NSGA-II employs a fast non-dominated sorting based on Pareto optimality~\cite{DBLP:conf/icec/TamakiKK96} and crowding-distance based comparison.
	      Such an approach will produce solutions that offer the best trade-off between competitive objectives~\cite{DBLP:conf/icec/TamakiKK96}.
\end{itemize}

\begin{figure}[t]
	\centering
	\includegraphics{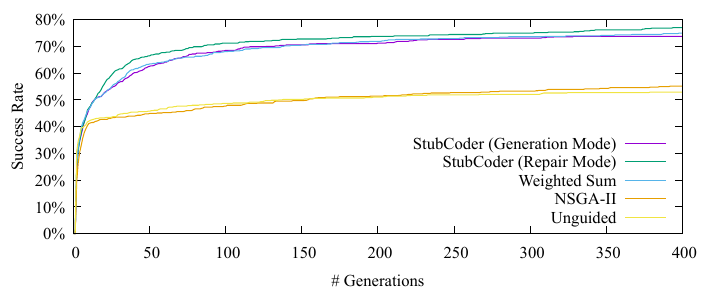}
	\vspace{-1em}
	\caption*{\footnotesize The value of Y-axis is the average success rate over the 10 repetitions at the generation budget specified by the X-axis.}
	\vspace{-1em}
	\caption{Comparison of the Success Rate between Experiment Setups}\label{fig:opt-comparison}
\end{figure}

\paragraph{Results.}
Column ``\tool{} (G)'' of Table~\ref{tab:results} shows the results for each test case.
Column ``SR'' shows the number of successful runs for each test case.
For successful runs, we also report the generations taken, time taken (in seconds), and the size of stub code, in columns ``Gen'', ``Time'', and ``\(|S|\)'', respectively.

For 45 of \numOfBenchmarkEntries{} test cases, \tool{} successfully generated test-passing stub code in at least 5 of the 10 repetitions, which counts for 76\% of the subjects.
The median of time taken by all the successful syntheses is 182 seconds.

Column \(\mid S\mid\) shows the length of the stub code (in terms of lines of code).
\tool{} is able to synthesize non-trivial stub code.
The type of variables in the stub code contains primitive types, strings, and complex objects.
For the 48 subjects with at least one successful run, 30 of them contain complex objects in the synthesized stub code.
For most of the test cases, the length of the stub code synthesized by \tool{} is slightly longer than the ones written by developers.
This is because \tool{} does not inline the variables that are used only once in the synthesized stub code, which is often done by developers \eg{In Listing~\ref{lst:benchmark-36}, developers put the String literal right in the \code{thenReturn}}.
Such a refactoring can be done trivially by using some refactoring tools.
For some test cases~\eg{\#2, \#30, and \#32}, \tool{} synthesized much fewer lines of code, yet obtained a test-passing stub code.
We found that for these test cases, developers copied the same stub code across several test cases, creating redundant stub calls.
Such a bad practice can complicate future maintenance of the stub code.
In comparison, the stub code synthesized by \tool{} will be easier to maintain.

\tool{} is more likely to be successful for subjects with simple stub code.
For instance, the stub codes in project MB3 mostly comprise a single line and do not contain complicated string values.
\tool{} is more likely to fail when a subject needs a complicated stub code to pass.
For example, in project JIB, there are multiple stub calls in the developer-written test containing complicated string values.
Although our fitness component \(AS\) can capture the edit distance between the expected value and the actual value in the assertions,
it cannot help when the string value returned by the stub code does not flow directly to the assertions \eg{used as branch conditions}.
Also, for project ZKN, the developer-written stub code contains a custom implementation that mutates the variables outside the stub, which is beyond the capability of \tool{}.

Figure~\ref{fig:opt-comparison} shows a comparison of the average success rate versus generation budget between different optimization strategies.
The dominance-based approach adopted by \tool{} performs similarly as weighted sum while it requires less parameter tuning effort.
NSGA-II performs much poorer than dominance-based approach and weighted sum.
The reason for this performance drop is that, NSGA-II aims to produce solutions that offer the best trade-off between competitive objectives~\cite{DBLP:conf/icec/TamakiKK96}, which will treat all the optimization objectives as of equal importance.
However, in our scenario, the three objectives measures the quality of candidate stub code in different stages of test execution and they are of different importance.
In this case, we can observe a large performance gap between NSGA-II and dominance-based approach.

\begin{answertorq}
	\tool{} successfully synthesizes the stub code for 76\% of the test cases in our benchmark in at least half of the repetitions.
	Optimization strategies that consider the importance of each objectives help \tool{} achieve better performance.
\end{answertorq}

\subsection{RQ2: Stub Code Repair}
\paragraph{Experiment Setup.}
The application scenario of RQ2 is stub code repair.
To answer RQ2, we followed the same setup as in RQ1, with the only difference that we fed \tool{} with the broken stub code so that it could make use of its tokens to construct the symbol pool.

\tool{} aims to generate and repair stub code in unit tests.
It is the first technique of its kind.
Existing test script repair techniques either repair oracle assertions~\cite{daniel2009reassert,daniel2010test}, repair GUI test scripts~\cite{ choudhary2011water,gao2015sitar,stocco2018visual}, or focus only on CUT calls~\cite{li2019intent,mirzaaghaei2012supporting}.
These techniques cannot repair obsolete stub code.
As a result, we do not use them as baselines in our evaluation.
Instead, we selected state-of-the-art program repair techniques as our baselines.
Specifically, we selected the techniques using the following criteria:
\begin{itemize}
	\item It is most recently published at a peer-reviewed venue.
	\item It has an artifact that works for \java{} projects.
	\item It needs only faulty code and test cases as input.
\end{itemize}
Following these criteria, we selected two state-of-the-art program repair techniques: \arja{}~\cite{DBLP:journals/tse/YuanB20} and the \cardumen{} mode~\cite{DBLP:conf/ssbse/MartinezM18} of \astor{}, and we applied the two baselines on our subjects.
Since the stub code being repaired is in the form of test code, which is not supported by these baselines,
we make the following adaptations for our subjects.
\begin{itemize}
	\item
	      Since the two baselines can only repair application code, we migrate the test case that contains the broken stub code (together with its dependencies) to the application code directory.
	\item
	      The baselines rely on fault localization techniques to find repair candidates.
	      However, in our scenario, the statements to be repaired are already known.
	      Therefore, we implement a fault localizer that returns the statements containing the obsolete stub code as faulty locations.
	      This can force the baselines techniques to repair only the stub code.
	\item
	      For each migrated test case, we create a simple test case to trigger it.
	      We specify these simple test cases as the failing tests when applying the two baselines.
\end{itemize}
After applying the adaptation, we run the two baselines with their default configurations with a time budget of six hours.

\paragraph{Results.}
As shown in Table~\ref{tab:results} (Column~``\tool{} (R)''), \tool{} successfully repaired 76\% of the test cases in our benchmark in no fewer than 5 repetitions.
There are 30 subjects where \tool{} synthesized stub code with complex objects.
As shown in Figure~\ref{fig:opt-comparison}, \tool{} took fewer generations to find a test-passing stub code.
With the help of the tokens in the broken stub code, \tool{} is able to produce test-passing stub code for the subjects where it fails in generation mode.
Take subject \#36 as an example, a complicated string \code{"/actuator/health"} must be stubbed to pass the test.
During code evolution, the signature of the method being stubbed changed, and the stub code was broken.
Nevertheless, the string literal in the broken stub code is still useful for \tool{} and enables it to converge quickly to the stub code that makes the test pass.
As shown in Listing~\ref{lst:benchmark-36rep}, \tool{} synthesized a two-line stub code to pass the test by reusing the literal string in the broken stub code, which was done in only four generations.
In comparison, without the help of the tokens, it is hard for \tool{} to synthesize such a complicated string literal from scratch and therefore, \tool{} failed to synthesize a test-passing stub code in generations mode.

For the two state-of-the-art baseline techniques, they failed to repair the stub code in any of our evaluation subjects.
There are two reasons for the poor performance achieved by the baseline techniques.
First, for 35 of the 59 subjects, the broken stub code leads to a compilation error.
The baseline techniques require compiling tests to run and therefore, are not applicable to these subjects.
Second, for the remaining 24 subjects, they failed to repair the stub code because they lack awareness of the semantics of the APIs in the mocking frameworks (mocking APIs).
Without understanding the mocking APIs, it is difficult for such techniques to find a test-passing stub code by randomly mutating the AST nodes.

\begin{answertorq}
	\tool{} successfully repairs the stub code for 76\% of the test cases in our benchmark in no fewer than 5 repetitions.
	The tokens in the broken stub code can help reduce search effort and synthesize shorter stub code in some cases.
	State-of-the-art program repair techniques cannot repair the stub code in any of the evaluation subjects.
\end{answertorq}

\subsection{RQ3: Effectiveness of Fitness Function}

\paragraph{Experiment Setup.}
RQ3 aims to evaluate the contribution of the fitness function to steering the search for stub code.
Towards this goal, we constructed a variant of \tool{} with random selection, which conducts the search process without the guidance of the fitness function.
Enumerating and (uniformly) sampling the whole search space would have been the ideal random baseline.
However, it is infeasible due to the huge size of the search space.
As such, we opted for a variant of \tool{} that uses the same crossover and mutation operations to explore the search space, but without any guidance by the fitness function.
We ran this variant of \tool{} with the same configurations as in RQ1.

\paragraph{Results.}
Column ``Unguided'' of Table~\ref{tab:results} shows the performance of the unguided variant of \tool.
Without the support of the fitness function, the unguided variant only successfully synthesizes stub code for 54\% of the test cases in our benchmark in no fewer than 5 repetitions, which is less than the generation mode and the repair mode.
In general, when the unguided variant successfully synthesizes stub code, it takes significantly more generations~\eg{\#4, \#28, and \#56} than the guided version of \tool.
For five test cases, only the unguided variant of \tool{} fails to synthesize the stub code \eg{\#30 and \#33}.
Interestingly, four of them have multiple oracle assertions in their test oracle.
For such test cases, \tool{} with fitness guidance can successfully synthesize the stub code because the fitness function examines the status of each assertion in the test oracle, and thus can prioritize the candidate stub code that can satisfy some of the assertions.
Such results show that our fitness function can effectively guide the search for test-passing stub code.

\begin{answertorq}
	\tool{} outperforms its unguided variant in both the generation and repair modes.
	Our fitness function provides useful guidance for synthesis of stub codes.
\end{answertorq}

\subsection{RQ4: Fidelity of Synthesized Stub Code}\label{sec:rq4}

\paragraph{Experiment setup.}
RQ1 and RQ2 evaluate the effectiveness of \tool{} in generating and repairing stub code that makes the developer-specified assertions pass.
Different from them, RQ4 evaluates the fidelity of the stub code synthesized by \tool{} with respect to the ground-truth stub code.
Specifically, we evaluate to what extent the test cases with synthesized stub code can preserve the runtime behavior of the test case with ground-truth stub code.
For each of the test cases in which \tool{} successfully synthesizes stub code in at least one run, we prepared \(\tau_{gt}=\langle V, S_{gt}, E, A\rangle\) with the stub code written by developers, and \(\tau_s = \langle V, S, E, A \rangle\) with the stub code synthesized by \tool{}.
Next, we opted for the similarities in three metrics to estimate similarity in the runtime behaviors of \(\tau_s\) and \(\tau_{gt}\).
A higher similarity in the runtime behaviors indicates a higher fidelity of the synthesized stub code.
\begin{itemize}
	\item \textbf{Executed Instructions.}
	      This metric measures the behavior of the test case with respect to exercising the code under test.
	      In this paper, we identify the set of \java{} bytecode instructions in the production code that are executed by \(\tau_s\) and \(\tau_{gt}\), denoted as \(I(\tau_s)\) and \(I(\tau_{gt})\), respectively.
	      Test cases with similar runtime behaviors should execute similar sets of instructions.
	      Therefore, we also report the Jaccard similarity coefficient~\cite{jaccard} between \(I(\tau_s)\) and \(I(\tau_{gt})\).
	      However, similar sets of executed instructions are not our only metric, since it is not a sufficient condition for similar behaviors.
	      It is possible that two test cases behaving differently share similar sets of executed instructions.
	\item \textbf{Execution Path.}
	      In addition to the set of executed instructions, we also traced the execution paths, which are the ordered sequence of instructions that are executed by the test cases.
	      Comparing the execution paths of \(\tau_{gt}\) and \(\tau_{s}\) would give us more information about fidelity because, unlike executed instructions, the execution path captures the instruction execution order.
	      For each of the test cases where \tool{} successfully synthesizes stub code, we collected the execution paths generated by \(\tau_s\) and \(\tau_{gt}\), denoted as \(P(\tau_s)\) and \(P(\tau_{gt})\), respectively.
	      Since Jaccard similarity coefficient cannot be applied to execution paths, we report their similarity based on edit distances as follows.
	      \[
		      \text{Similarity}\left(P(\tau_s), P(\tau_{gt})\right) = 1 - \frac{DLev\left(P(\tau_s), P(\tau_{gt})\right)}{|P(\tau_s)| + |P(\tau_{gt})|}
	      \]
	      where \(DLev\) is the Damerau–Levenshtein distance~\cite{Damerau_Levenshtein_distance}.
	      For test cases spawning multiple threads, we match the threads that share similar traces, and \(DLev\) denotes the sum of Damerau-Levenshtein distances between those thread pairs.
	      A small edit distance indicates that \(\tau_s\) and \(\tau_{gt}\) traverse similar execution paths.

	\item \textbf{Killed Mutants.}
	      Mutation analysis~\cite{DBLP:journals/tse/JiaH11} measures the adequacy of a test case with respect to detecting faults.
          It injects artificial faults in the program and checks if the test cases can ``kill'' them \ie{the test fails}.
	      In this paper, we mutated the CUT using \textsc{PIT}~\cite{DBLP:conf/issta/ColesLHPV16} by seeding faults and ran \(\tau_s\) and \(\tau_{gt}\) against the mutants.
	      We identified the mutants that are killed by \(\tau_s\) and \(\tau_{gt}\), denoted as \(K(\tau_s)\) and \(K(\tau_{gt})\), respectively.
	      Test cases with similar behaviors should be able to kill similar sets of mutants.
	      Therefore, we also report the Jaccard similarity coefficient~\cite{jaccard} between \(K(\tau_s)\) and \(K(\tau_{gt})\).
\end{itemize}

We choose these metrics because they estimate the intent or behaviors of test cases.
For example, the executed instructions and execution path are relaxed and tighten versions of path conditions.
They are validated to be a good abstraction of test intents in a recent study on test repair~\cite{li2019intent}.
Mutation coverage is a proxy for reflecting the behaviors of the test cases in terms of detecting potential bugs, and it was used to measure the behavioral similarity of test cases in a recent study that automatically refactors test cases with mocking~\cite{DBLP:conf/sigsoft/WangXYWW21}.

\begin{table}[]
	\caption{RQ4: Fidelity of the Synthesized Stub Code (Generation Mode)}\label{tab:fidelity-generation}
	\vspace{-0.6em}
	\captionsetup{style=base,singlelinecheck=off,font=scriptsize}
	\caption*{
		\(\tau_{gt}\) denotes the result generated by the ground truth.
		\(\tau_{s}\) denotes the result generated by the test with synthesized stub code.\\
		\(\tau_{gt}\cap\tau_s\) denotes the intersection of the ground truth and the test with synthesized stub code.\\
		Jaccard denotes Jaccard similarity coefficient.\\
		\(DLev\) denotes Damerau-Levenshtein distance.
		\smallskip{}
	}
	\vspace{-1em}
	\scriptsize
	\renewcommand{\arraystretch}{0.9}
	\begin{center}
    \setlength{\tabcolsep}{4.5pt}
    \begin{tabularx}{\linewidth}{ll|rrrr|rrrr|rrrrr}
        \toprule
        \multicolumn{2}{l}{\multirow{3}{*}{\textbf{Subject ID}}}
        &\multicolumn{4}{c}{\textbf{Executed Instructions}} 
        & \multicolumn{4}{c}{\textbf{Execution Path}}
        & \multicolumn{5}{c}{\textbf{Killed Mutants}}
        \\
        \cmidrule(l{1pt}r{1pt}){3-6}
        \cmidrule(l{1pt}r{1pt}){7-10}
        \cmidrule(l{1pt}r{1pt}){11-15}
        &&
        \(\tau_{gt}\) & \(\tau_s\) & \(\tau_{gt}\cap \tau_s\) & Jaccard &
        \(\tau_{gt}\) & \(\tau_s\) & \(DLev\)                 & Similarity &
        Injected & \(\tau_{gt}\) & \(\tau_s\) & \(\tau_{gt}\cap \tau_s\) & Jaccard \\
        \midrule[\heavyrulewidth]
        \multirow{5}{*}{N4J}  & \#1    & 40   &  40   & 40   & 100.00\%      & 40    & 40    & 0    & 100.00\%  & 27  & 3  & 3  & 3  & 100.00\%  \\
                              & \#2    & 431  &  431  & 431  & 100.00\%      & 564   & 561   & 3    & 99.73\%   & 32  & 4  & 3  & 3  & 75.00\%   \\
                              & \#3    & -    &  -    & -    & -             & -     & -     & -    & -         & -   & -  & -  & -  & -         \\
                              & \#4    & 258  &  258  & 258  & 100.00\%      & 258   & 258   & 0    & 100.00\%  & 378 & 12 & 11 & 11 & 91.67\%   \\
\midrule                SPB   & \#5    & 117  &  117  & 117  & 100.00\%      & 171   & 171   & 0    & 100.00\%  & 2   & 1  & 1  & 1  & 100.00\%  \\
\midrule\multirow{3}{*}{GRC}  & \#6    & 255  &  255  & 255  & 100.00\%      & 257   & 257   & 0    & 100.00\%  & 89  & 2  & 2  & 2  & 100.00\%  \\
                              & \#7    & -    &  -    & -    & -             & -     & -     & -    & -         & -   & -  & -  & -  & -         \\
                              & \#8    & -    &  -    & -    & -             & -     & -     & -    & -         & -   & -  & -  & -  & -         \\
\midrule\multirow{12}{*}{MB3} & \#9    & 39   &  39   & 39   & 100.00\%      & 39    & 39    & 0    & 100.00\%  & 5   & 1  & 1  & 1  & 100.00\%  \\
                              & \#10   & 39   &  39   & 39   & 100.00\%      & 39    & 39    & 0    & 100.00\%  & 5   & 1  & 1  & 1  & 100.00\%  \\
                              & \#11   & 39   &  39   & 39   & 100.00\%      & 39    & 39    & 0    & 100.00\%  & 5   & 1  & 1  & 1  & 100.00\%  \\
                              & \#12   & 39   &  39   & 39   & 100.00\%      & 39    & 39    & 0    & 100.00\%  & 5   & 1  & 1  & 1  & 100.00\%  \\
                              & \#13   & 39   &  39   & 39   & 100.00\%      & 39    & 39    & 0    & 100.00\%  & 5   & 1  & 1  & 1  & 100.00\%  \\
                              & \#14   & 39   &  39   & 39   & 100.00\%      & 39    & 39    & 0    & 100.00\%  & 5   & 1  & 1  & 1  & 100.00\%  \\
                              & \#15   & 39   &  39   & 39   & 100.00\%      & 39    & 39    & 0    & 100.00\%  & 5   & 1  & 1  & 1  & 100.00\%  \\
                              & \#16   & 39   &  39   & 39   & 100.00\%      & 39    & 39    & 0    & 100.00\%  & 5   & 1  & 1  & 1  & 100.00\%  \\
                              & \#17   & 39   &  39   & 39   & 100.00\%      & 39    & 39    & 0    & 100.00\%  & 5   & 1  & 1  & 1  & 100.00\%  \\
                              & \#18   & 39   &  39   & 39   & 100.00\%      & 39    & 39    & 0    & 100.00\%  & 5   & 1  & 1  & 1  & 100.00\%  \\
                              & \#19   & 39   &  39   & 39   & 100.00\%      & 39    & 39    & 0    & 100.00\%  & 5   & 1  & 1  & 1  & 100.00\%  \\
                              & \#20   & 39   &  39   & 39   & 100.00\%      & 39    & 39    & 0    & 100.00\%  & 5   & 1  & 1  & 1  & 100.00\%  \\
\midrule\multirow{5}{*}{ZKN}  & \#21   & 55   &  41   & 41   & 74.55\%       & 57    & 41    & 16   & 83.67\%   & 29  & 4  & 1  & 1  & 25.00\%   \\
                              & \#22   & -    &  -    & -    & -             & -     & -     & -    & -         & -   & -  & -  & -  & -         \\
                              & \#23   & 46   &  4    & 4    & 8.7\%         & 46    & 4     & 42   & 16.00\%   & 29  & 2  & 0  & 0  & 0.00\%    \\
                              & \#24   & -    &  -    & -    & -             & -     & -     & -    & -         & -   & -  & -  & -  & -         \\
                              & \#25   & 46   &  4    & 4    & 8.7\%         & 62    & 4     & 58   & 12.12\%   & 29  & 3  & 0  & 0  & 0.00\%    \\
\midrule\multirow{8}{*}{APL}  & \#26   & 40   &  40   & 40   & 100.00\%      & 40    & 40    & 0    & 100.00\%  & 5   & 0  & 3  & 0  & 0.00\%    \\
                              & \#27   & 28   &  28   & 28   & 100.00\%      & 28    & 28    & 0    & 100.00\%  & 5   & 2  & 2  & 2  & 100.00\%  \\
                              & \#28   & 327  &  317  & 317  & 96.94\%       & 336   & 324   & 12   & 98.18\%   & 40  & 8  & 7  & 7  & 87.50\%   \\
                              & \#29   & 118  &  99   & 99   & 83.90\%       & 120   & 99    & 21   & 90.41\%   & 22  & 8  & 8  & 6  & 60.00\%   \\
                              & \#30   & 265  &  265  & 265  & 100.00\%      & 396   & 396   & 0    & 100.00\%  & 40  & 6  & 6  & 6  & 100.00\%  \\
                              & \#31   & -    &  -    & -    & -             & -     & -     & -    & -         & -   & -  & -  & -  & -         \\
                              & \#32   & 113  &  35   & 35   & 30.97\%       & 113   & 35    & 78   & 47.30\%   & 22  & 10 & 3  & 2  & 18.18\%   \\
                              & \#33   & -    &  -    & -    & -             & -     & -     & -    & -         & -   & -  & -  & -  & -         \\
\midrule                ZUL   & \#34   & 31   &  31   & 31   & 100.00\%      & 31    & 31    & 0    & 100.00\%  & 4   & 2  & 2  & 2  & 100.00\%  \\
\midrule\multirow{2}{*}{SBA}  & \#35   & -    &  -    & -    & -             & -     & -     & -    & -         & -   & -  & -  & -  & -         \\
                              & \#36   & -    &  -    & -    & -             & -     & -     & -    & -         & -   & -  & -  & -  & -         \\
\midrule\multirow{2}{*}{JIB}  & \#37   & -    &  -    & -    & -             & -     & -     & -    & -         & -   & -  & -  & -  & -         \\
                              & \#38   & -    &  -    & -    & -             & -     & -     & -    & -         & -   & -  & -  & -  & -         \\
\midrule                AZK   & \#39   & -    &  -    & -    & -             & -     & -     & -    & -         & -   & -  & -  & -  & -         \\
\midrule\multirow{5}{*}{AHP}  & \#40   & 2197 &  2027 & 2026 & 92.17\%       & 5083  & 4841  & 243  & 97.55\%   & 288 & 8  & 5  & 5  & 62.50\%   \\
                              & \#41   & 903  &  903  & 903  & 100.00\%      & 9995  & 9998  & 5    & 99.97\%   & 142 & 15 & 15 & 15 & 100.00\%  \\
                              & \#42   & 1500 &  1436 & 1432 & 95.21\%       & 11144 & 11016 & 136  & 99.39\%   & 147 & 17 & 18 & 17 & 94.44\%   \\
                              & \#43   & 1055 &  1055 & 1055 & 100.00\%      & 9343  & 9346  & 5    & 99.97\%   & 120 & 11 & 17 & 11 & 64.71\%   \\
                              & \#44   & 1056 &  1093 & 1054 & 96.26\%       & 9217  & 5400  & 3964 & 72.88\%   & 120 & 14 & 15 & 14 & 93.33\%   \\
\midrule\multirow{2}{*}{ADR}  & \#45   & 59   &  59   & 59   & 100.00\%      & 67    & 67    & 0    & 100.00\%  & 17  & 3  & 3  & 3  & 100.00\%  \\
                              & \#46   & 169  &  164  & 150  & 81.97\%       & 320   & 208   & 126  & 76.14\%   & 29  & 4  & 4  & 2  & 33.33\%   \\
\midrule\multirow{13}{*}{ADU} & \#47   & 188  &  174  & 173  & 91.53\%       & 235   & 221   & 15   & 96.71\%   & 5   & 1  & 1  & 1  & 100.00\%  \\
                              & \#48   & 61   &  62   & 60   & 95.24\%       & 66    & 67    & 2    & 98.50\%   & 30  & 5  & 5  & 5  & 100.00\%  \\
                              & \#49   & 74   &  60   & 60   & 81.08\%       & 78    & 62    & 16   & 88.57\%   & 22  & 2  & 2  & 2  & 100.00\%  \\
                              & \#50   & 70   &  56   & 56   & 80.00\%       & 74    & 58    & 16   & 87.88\%   & 22  & 1  & 2  & 1  & 50.00\%   \\
                              & \#51   & 41   &  12   & 12   & 29.27\%       & 41    & 12    & 29   & 45.28\%   & 5   & 1  & 2  & 1  & 50.00\%   \\
                              & \#52   & 64   &  64   & 64   & 100.00\%      & 66    & 66    & 0    & 100.00\%  & 2   & 1  & 1  & 1  & 100.00\%  \\
                              & \#53   & 356  &  100  & 88   & 23.91\%       & 448   & 106   & 355  & 35.92\%   & 29  & 2  & 1  & 0  & 0.00\%    \\
                              & \#54   & -    &  -    & -    & -             & -     & -     & -    & -         & -   & -  & -  & -  & -         \\
                              & \#55   & 64   &  64   & 64   & 100.00\%      & 66    & 66    & 0    & 100.00\%  & 2   & 1  & 1  & 1  & 100.00\%  \\
                              & \#56   & 387  &  100  & 88   & 22.06\%       & 465   & 106   & 372  & 34.85\%   & 29  & 2  & 2  & 0  & 0.00\%    \\
                              & \#57   & 296  &  178  & 165  & 53.40\%       & 372   & 194   & 247  & 56.36\%   & 22  & 1  & 2  & 0  & 0.00\%    \\
                              & \#58   & 155  &  143  & 143  & 92.26\%       & 185   & 173   & 12   & 96.65\%   & 29  & 8  & 8  & 8  & 100.00\%  \\
                              & \#59   & 92   &  6    & 6    & 6.52\%        & 95    & 6     & 90   & 10.89\%   & 5   & 2  & 0  & 0  & 0.00\%    \\
        \midrule[\heavyrulewidth]
        \multicolumn{2}{l}{\textbf{Median}}&&&               & 100.00\%&              &&            & 99.99\%   &&&&            & 100.00\%   \\ 
        \bottomrule
    \end{tabularx}
\end{center}

\end{table}

\begin{table}[]
	\caption{RQ4: Fidelity of the Synthesized Stub Code (Repair Mode)}\label{tab:fidelity-repair}
	\vspace{-0.6em}
	\captionsetup{style=base,singlelinecheck=off,font=scriptsize}
	\caption*{
		\(\tau_{gt}\) denotes the result generated by the ground truth.
		\(\tau_{s}\) denotes the result generated by the test with synthesized stub code.\\
		\(\tau_{gt}\cap\tau_s\) denotes the intersection of the ground truth and the test with synthesized stub code.\\
		Jaccard denotes Jaccard similarity coefficient.\\
		\(DLev\) denotes Damerau-Levenshtein distance.
		\smallskip{}
	}
	\vspace{-1em}
	\scriptsize
	\renewcommand{\arraystretch}{0.9}
	
\begin{center}
    \setlength{\tabcolsep}{4.5pt}
    \begin{tabularx}{\linewidth}{ll|rrrr|rrrr|rrrrr}
        \toprule
        \multicolumn{2}{l}{\multirow{3}{*}{\textbf{Subject ID}}}
        &\multicolumn{4}{c}{\textbf{Executed Instructions}} 
        & \multicolumn{4}{c}{\textbf{Execution Path}}
        & \multicolumn{5}{c}{\textbf{Killed Mutants}}
        \\
        \cmidrule(l{1pt}r{1pt}){3-6}
        \cmidrule(l{1pt}r{1pt}){7-10}
        \cmidrule(l{1pt}r{1pt}){11-15}
        &&
        \(\tau_{gt}\) & \(\tau_s\) & \(\tau_{gt}\cap \tau_s\) & Jaccard &
        \(\tau_{gt}\) & \(\tau_s\) & \(DLev\)                 & Similarity &
        Injected & \(\tau_{gt}\) & \(\tau_s\) & \(\tau_{gt}\cap \tau_s\) & Jaccard \\
        \midrule[\heavyrulewidth]
        \multirow{4}{*}{N4J}  & \#1    & 40   & 40   & 40   & 100.00\%     & 40     & 40    & 0    & 100.00\%    & 27  & 3  & 3  & 3  & 100.00\%  \\
                              & \#2    & 431  & 431  & 431  & 100.00\%     & 564    & 559   & 5    & 99.55\%     & 32  & 4  & 3  & 3  & 75.00\%   \\
                              & \#3    & 305  & 277  & 277  & 90.82\%      & 327    & 293   & 34   & 94.52\%     & 67  & 15 & 14 & 13 & 81.25\%   \\
                              & \#4    & 258  & 385  & 258  & 67.01\%      & 258    & 389   & 131  & 79.75\%     & 378 & 11 & 12 & 11 & 91.67\%   \\
\midrule                SPB   & \#5    & 117  & 117  & 117  & 100.00\%     & 171    & 171   & 0    & 100.00\%    & 2   & 1  & 1  & 1  & 100.00\%  \\
\midrule\multirow{3}{*}{GRC}  & \#6    & 255  & 255  & 255  & 100.00\%     & 257    & 257   & 0    & 100.00\%    & 89  & 2  & 2  & 2  & 100.00\%  \\
                              & \#7    & -    & -    & -    & -            & -      & -     & -    & -           & -   & -  & -  & -  & -         \\
                              & \#8    & -    & -    & -    & -            & -      & -     & -    & -           & -   & -  & -  & -  & -         \\
\midrule\multirow{12}{*}{MB3} & \#9    & 39   & 39   & 39   & 100.00\%     & 39     & 39    & 0    & 100.00\%    & 5   & 1  & 1  & 1  & 100.00\%  \\
                              & \#10   & 39   & 39   & 39   & 100.00\%     & 39     & 39    & 0    & 100.00\%    & 5   & 1  & 1  & 1  & 100.00\%  \\
                              & \#11   & 39   & 39   & 39   & 100.00\%     & 39     & 39    & 0    & 100.00\%    & 5   & 1  & 1  & 1  & 100.00\%  \\
                              & \#12   & 39   & 39   & 39   & 100.00\%     & 39     & 39    & 0    & 100.00\%    & 5   & 1  & 1  & 1  & 100.00\%  \\
                              & \#13   & 39   & 39   & 39   & 100.00\%     & 39     & 39    & 0    & 100.00\%    & 5   & 1  & 1  & 1  & 100.00\%  \\
                              & \#14   & 39   & 39   & 39   & 100.00\%     & 39     & 39    & 0    & 100.00\%    & 5   & 1  & 1  & 1  & 100.00\%  \\
                              & \#15   & 39   & 39   & 39   & 100.00\%     & 39     & 39    & 0    & 100.00\%    & 5   & 1  & 1  & 1  & 100.00\%  \\
                              & \#16   & 39   & 39   & 39   & 100.00\%     & 39     & 39    & 0    & 100.00\%    & 5   & 1  & 1  & 1  & 100.00\%  \\
                              & \#17   & 39   & 39   & 39   & 100.00\%     & 39     & 39    & 0    & 100.00\%    & 5   & 1  & 1  & 1  & 100.00\%  \\
                              & \#18   & 39   & 39   & 39   & 100.00\%     & 39     & 39    & 0    & 100.00\%    & 5   & 1  & 1  & 1  & 100.00\%  \\
                              & \#19   & 39   & 39   & 39   & 100.00\%     & 39     & 39    & 0    & 100.00\%    & 5   & 1  & 1  & 1  & 100.00\%  \\
                              & \#20   & 39   & 39   & 39   & 100.00\%     & 39     & 39    & 0    & 100.00\%    & 5   & 1  & 1  & 1  & 100.00\%  \\
\midrule\multirow{5}{*}{ZKN}  & \#21   & 55   & 41   & 41   & 74.55\%      & 57     & 41    & 16   & 83.67\%     & 29  & 4  & 1  & 1  & 25.00\%   \\
                              & \#22   & -    & -    & -    & -            & -      & -     & -    & -           & -   & -  & -  & -  & -         \\
                              & \#23   & 46   & 4    & 4    & 8.70\%       & 46     & 4     & 42   & 16.00\%     & 29  & 2  & 0  & 0  & 0.00\%    \\
                              & \#24   & -    & -    & -    & -            & -      & -     & -    & -           & -   & -  & -  & -  & -         \\
                              & \#25   & 46   & 4    & 4    & 8.70\%       & 62     & 4     & 58   & 12.12\%     & 29  & 3  & 0  & 0  & 0.00\%    \\
\midrule\multirow{8}{*}{APL}  & \#26   & 40   & 40   & 40   & 100.00\%     & 40     & 40    & 0    & 100.00\%    & 5   & 0  & 3  & 0  & 0.00\%    \\
                              & \#27   & 28   & 111  & 28   & 25.23\%      & 28     & 111   & 83   & 40.29\%     & 5   & 2  & 2  & 2  & 100.00\%  \\
                              & \#28   & 327  & 317  & 317  & 96.94\%      & 336    & 324   & 12   & 98.18\%     & 40  & 8  & 7  & 7  & 87.50\%   \\
                              & \#29   & 118  & 99   & 99   & 83.90\%      & 120    & 99    & 21   & 90.41\%     & 22  & 9  & 9  & 7  & 63.64\%   \\
                              & \#30   & 265  & 265  & 265  & 100.00\%     & 396    & 396   & 0    & 100.00\%    & 40  & 6  & 6  & 6  & 100.00\%  \\
                              & \#31   & -    & -    & -    & -            & -      & -     & -    & -           & -   & -  & -  & -  & -         \\
                              & \#32   & 113  & 35   & 35   & 30.97\%      & 113    & 35    & 78   & 47.30\%     & 22  & 6  & 3  & 1  & 12.50\%   \\
                              & \#33   & -    & -    & -    & -            & -      & -     & -    & -           & -   & -  & -  & -  & -         \\
\midrule                ZUL   & \#34   & 31   & 31   & 31   & 100.00\%     & 31     & 31    & 0    & 100.00\%    & 4   & 2  & 2  & 2  & 100.00\%  \\
\midrule\multirow{2}{*}{SBA}  & \#35   & 268  & 268  & 268  & 100.00\%     & 279    & 279   & 0    & 100.00\%    & 49  & 15 & 14 & 14 & 93.33\%   \\
                              & \#36   & 297  & 297  & 297  & 100.00\%     & 374    & 374   & 0    & 100.00\%    & 4   & 2  & 2  & 2  & 100.00\%  \\
\midrule\multirow{2}{*}{JIB}  & \#37   & -    & -    & -    & -            & -      & -     & -    & -           & -   & -  & -  & -  & -         \\
                              & \#38   & 1245 & 1228 & 1228 & 98.63\%      & 1588   & 1465  & 123  & 95.97\%     & 46  & 12 & 12 & 12 & 100.00\%  \\
\midrule                AZK   & \#39   & -    & -    & -    & -            & -      & -     & -    & -           & -   & -  & -  & -  & -         \\
\midrule\multirow{5}{*}{AHP}  & \#40   & 2197 & 2027 & 2026 & 92.17\%      & 5083   & 4841  & 243  & 97.55\%     & 288 & 8  & 5  & 5  & 62.50\%   \\
                              & \#41   & 903  & 901  & 901  & 99.78\%      & 9995   & 5460  & 4535 & 70.66\%     & 142 & 15 & 15 & 15 & 100.00\%  \\
                              & \#42   & 1500 & 1436 & 1432 & 95.21\%      & 11144  & 11016 & 136  & 99.39\%     & 147 & 17 & 18 & 17 & 94.44\%   \\
                              & \#43   & 1055 & 1094 & 1055 & 96.44\%      & 9343   & 9526  & 183  & 99.03\%     & 120 & 11 & 12 & 11 & 91.67\%   \\
                              & \#44   & 1056 & 1093 & 1054 & 96.26\%      & 9217   & 5400  & 3964 & 72.88\%     & 120 & 14 & 15 & 14 & 93.33\%   \\
\midrule\multirow{2}{*}{ADR}  & \#45   & -    & -    & -    & -            & -      & -     & -    & -           & -   & -  & -  & -  & -         \\
                              & \#46   & 169  & 164  & 150  & 81.97\%      & 320    & 208   & 126  & 76.14\%     & 29  & 4  & 4  & 2  & 33.33\%   \\
\midrule\multirow{13}{*}{ADU} & \#47   & 188  & 174  & 173  & 91.53\%      & 235    & 221   & 15   & 96.71\%     & 5   & 1  & 1  & 1  & 100.00\%  \\
                              & \#48   & 61   & 61   & 61   & 100.00\%     & 66     & 66    & 0    & 100.00\%    & 30  & 5  & 5  & 5  & 100.00\%  \\
                              & \#49   & 74   & 60   & 60   & 81.08\%      & 78     & 62    & 16   & 88.57\%     & 22  & 2  & 2  & 2  & 100.00\%  \\
                              & \#50   & 70   & 63   & 56   & 72.73\%      & 74     & 65    & 16   & 88.49\%     & 22  & 1  & 2  & 1  & 50.00\%   \\
                              & \#51   & 41   & 47   & 12   & 15.79\%      & 41     & 47    & 44   & 50.00\%     & 5   & 1  & 2  & 1  & 50.00\%   \\
                              & \#52   & 64   & 64   & 64   & 100.00\%     & 66     & 66    & 0    & 100.00\%    & 2   & 1  & 1  & 1  & 100.00\%  \\
                              & \#53   & 356  & 115  & 94   & 24.93\%      & 448    & 121   & 350  & 38.49\%     & 29  & 2  & 1  & 0  & 0.00\%    \\
                              & \#54   & -    & -    & -    & -            & -      & -     & -    & -           & -   & -  & -  & -  & -         \\
                              & \#55   & 64   & 64   & 64   & 100.00\%     & 66     & 66    & 0    & 100.00\%    & 2   & 1  & 1  & 1  & 100.00\%  \\
                              & \#56   & 387  & 117  & 101  & 25.06\%      & 465    & 142   & 372  & 38.71\%     & 29  & 2  & 1  & 0  & 0.00\%    \\
                              & \#57   & 296  & 94   & 81   & 26.21\%      & 372    & 104   & 281  & 40.97\%     & 22  & 1  & 1  & 0  & 0.00\%    \\
                              & \#58   & 155  & 143  & 143  & 92.26\%      & 185    & 173   & 12   & 96.65\%     & 29  & 8  & 8  & 8  & 100.00\%  \\
                              & \#59   & 92   & 82   & 82   & 89.13\%      & 95     & 85    & 10   & 94.44\%     & 5   & 2  & 2  & 2  & 100.00\%  \\
        \midrule[\heavyrulewidth]
        \multicolumn{2}{l}{\textbf{Median}}&&&              & 99.78\%&             &&             & 99.39\%       &&&&               & 100.00\%    \\ 
        \bottomrule
    \end{tabularx}
\end{center}

\end{table}
\paragraph{Results.}
For generation mode, Table~\ref{tab:fidelity-generation} shows the comparisons on executed instructions, execution path, and killed mutants by \(\tau_s\) and \(\tau_{gt}\) for each subject in our benchmark that \tool{} successfully synthesizes stub code in at least one run.\footnote{PIT crashed due to its bug on subject \#54 and therefore we cannot report the result for \#54 in this RQ4.}
In our experiment, \(\tau_s\) covers the similar set of the instructions as \(\tau_{gt}\), with the median of the Jaccard similarity coefficient to be 100\%.
In 24 of the 46 subjects, \(\tau_s\) covers exactly the same set of instructions as \(\tau_{gt}\).
In such cases, \(\tau_s\) is capable for exercising the same instructions as \(\tau_{gt}\).
The execution paths traversed by \(\tau_s\) and \(\tau_{gt}\) are also similar.
The median of similarity is 99.99\%, which indicates that \(\tau_s\) exercise the CUT in a way similar to what \(\tau_{gt}\) does.
In 22 of the 46 subjects, the edit distance between the execution paths generated by \(\tau_s\) and \(\tau_{gt}\) is 0, which means that they execute the instructions in the production code in exactly the same order.
In such subjects, \(\tau_s\) and \(\tau_{gt}\) exercise the CUT with the same intent.
This is because the same execution path indicates that the tests share the same path conditions, which was shown to be a good abstraction of test intent~\cite{li2019intent}.
The set of mutants killed by \(\tau_s\) and \(\tau_{gt}\) are also similar, with the median of similarity to be 100\%.
In 26 of the 46 subjects, \(\tau_s\) kills exactly the same set of mutants as \(\tau_{gt}\), which means \(\tau_s\) has the similar ability to detect injected bugs as \(\tau_{gt}\).

Table~\ref{tab:fidelity-repair} gives the fidelity comparison for the repair more.
The median of similarity in executed instructions is 99.78\%, and \(\tau_s\) covers the same set of instructions as \(\tau_{gt}\) does in 24 of 49 subjects.
The median of similarity in execution path is 99.39\%, and \(\tau_s\) shares exactly the same execution path as \(\tau_{gt}\) in 23 of 49 subjects.
The median of similarity in killed mutants is 100\%, and \(\tau_s\) kills the same set of mutants as \(\tau_{gt}\) does in 28 of 49 subjects.
In summary, \tool{} synthesizes stub code similar to the ground truth with respect to executed instructions, execution path, and killed mutation.
The high similarities indicate that the test cases with the synthesized stub code have adequacy similar to that of the ground truth.
Such test cases are useful for detecting regression bugs when the CUT evolves.

Besides most of the high hifelities subjects, we also observed several cases that worth discussion:
\begin{itemize}
	\item \textbf{Synthesis is successful but fidelity is low.}
	      Such cases are caused by weak test oracles.
	      The test oracles in these subjects \eg{\#23 and \#25 in both modes} allow multiple execution paths to pass the test.
	      Listing~\ref{lst:weak-oracle-example} shows an example to illustrate such cases.
	      As shown in the code snippet, both \(\tau_{gt}\) and \(\tau\) can pass the test since both of them will result in a \code{CustomException} to be thrown.
	      However, the execution paths of \(\tau_{gt}\) and \(\tau\) are different, and the mutants injected into the branch cannot be killed by \(\tau\).
	      Such situations can be easily mitigated by enhancing the test oracle with a few mocking calls, specifying that certain methods should be called on the mock objects.
	      After that, \tool{} will synthesize stub code that helps cover the code lines invoking the methods specified by such mocking calls.
	\item \textbf{Killed mutants are the same but execution paths are different.}
	      Such cases \eg{\#47, \#49, and \#58 in both modes} happened because \(\tau_{gt}\) and \(\tau\) takes different ways to construct certain objects.
	      To illustrate such a difference, Listing~\ref{lst:alt-object-construction} show a comparison of two ways to construct a string \code{"10"}.
	      In the developer-written stub code, the string is constructed directly with a literal.
	      In the synthesized stub code, the string is converted from an interger value.
	      In this case, the instructions in \code{String.valueOf} will be included in the execution path of \(\tau\) but not \(\tau_{gt}\).
	      Nevertheless, the synthesized stub code is sill useful for developers as \(\tau\) kills exactly the same set of mutants as \(\tau_{gt}\).
	\item \textbf{Execution paths are similar but killed mutants are different.}
	      Such cases \eg{\#2, \#35 in repair mode} are due to different return values specified in the stub code.
	      Listing~\ref{lst:val-spec} illustrates such a case.
	      The generated stub code and the developer-written stub code specify different return values for the method \code{getOffset}.
	      Without mutation, both test cases can enter the then branch, and therefore they share the same set of executed instructions, and both of them make the test pass.
	      However, when the mutation operator changes the \code{+} to \code{-}, the test case with developer-written stub code does not enter the then branch while the test case with generated stub code enters the then branch.
	      In this case, the test fails with the developer-written code while it passes with the generated stub code.
	\item \textbf{Same executed instructions but different execution paths.}
	      This is because there are loops in the production code and \(\tau\) and \(\tau_{gt}\) executed the loops for different number of times.
\end{itemize}

\begin{figure}[t]
\begin{lstlisting}[
		language=java,
		morekeywords={var},
		caption={Illustration of Weak Oracle},
		label={lst:weak-oracle-example},
		escapechar=|,
		numbers=left
	]
// Synthesized stub code
when(someMock.getValue()).thenThrow(new CustomException());

// Developer-written stub code
when(someMock.getValue()).thenReturn(5);

// Class under test
int value = someMock.getValue();
if(value < 10) {
    // ... mutants are injected here
    throw CustomException();
}

// Assertions
assertThrows(CustomException.class, ()-> {
    // CUT execution
});
\end{lstlisting}
\end{figure}

\begin{figure}[t]
\begin{lstlisting}[
		language=java,
		morekeywords={var},
		caption={Illustration of Alternatve Object Construction},
		label={lst:alt-object-construction},
		escapechar=|,
		numbers=left
	]
// Generated code
int var1 = 10;
String var2 = String.valueOf(var1);
when(someMock.foo()).thenReturn(var2);

// Developer-written code
when(someMock.foo()).thenReturn("10");
\end{lstlisting}
\end{figure}

\begin{figure}[t]
\begin{lstlisting}[
		language=java,
		morekeywords={var},
		caption={Illustration of Different Killed Mutants due to Return Values},
		label={lst:val-spec},
		escapechar=|,
		numbers=left
	]
// Generated stub code
doReturn(2).when(someMock).getOffset();

// Developer-written stub code
doReturn(4).when(someMock).getOffset();

// In the class under test
int offset = someMock.getOffset() + 5;
if (3 <= offset && offset <= 10) {
    // ...
} else {
    // fail
}
\end{lstlisting}
\end{figure}

Overall, as shown in Table~\ref{tab:fidelity-generation} and Table~\ref{tab:fidelity-repair}, the encoded information is sufficient for obtaining a useful stub code most of the time.
This verifies our intuition that deriving the stub code from the information encoded in the CUT execution code and test oracle leads to adequate test cases.

\begin{answertorq}
	\tool{} synthesizes stub code with high fidelity, which means that they share a runtime behavior similar to that of the ground truth in terms of their effects on the code under test.
	The information encoded in CUT execution code and test oracle is useful for deriving stub code.
	\tool{} works well when the test oracle contains adequate information.
\end{answertorq}

\subsection{Threats to Validity}

\paragraph{Subject Collection.}
We evaluated \tool{} on \numOfBenchmarkEntries{} test cases collected from \numOfBenchmarkProjects{} projects.
Our results might not be generalized to other projects and test cases.
The subject collection requires intensive manual effort, which limited the number of projects and test cases that we could use.
To mitigate this threat, we selected
large, actively maintained, diverse, and popular GitHub projects.
These projects belong to different domains: big data, database, web apps, containers, etc.
Our benchmark dataset reflects the real-world usage of stub code in these areas.

Also, when preparing the evaluation subjects, we rewrote the assertions written in other libraries into those using \junit{} framework.
Such manual modification might be affected by human mistakes and thus change behavior of the test cases.
To mitigate this issue, we cross-checked the documentation of the corresponding assertion framework and \junit{} to make sure the rewritten assertions preserves the original semantics.
We also ran the test cases before and after modification to make sure that they yield the same result.

\paragraph{Fidelity Measurement.}
When measuring the fidelity of the stub code, we leveraged a metric based on instruction coverage.
However, the similarity on the instruction coverage may not be ideal to reflect the differences.
For example, when there are only 10\% of the instructions in the class under test are in branches, the similarity of the instruction coverage will be at least 90\%.
To mitigate this threat, we introduced additional metrics such a execution path and killed mutants to further characterize the behavior of the test cases.

\paragraph{Experiments.}
Evolutionary algorithms are stochastic by nature, and the evaluation results may be different across several runs.
In our experiments, we used 10 repetitions to evaluate \tool.
The effectiveness of \tool{} is likely to increase with more attempts and a higher budget.
However, \tool{} results are stable, as shown in Table~\ref{tab:results}.
There are only a few test cases where the 10 attempts gave inconsistent results.
Nevertheless, conducting more experiments is an important future work.

\section{Related Work}\label{sec:related-work}

\paragraph{Stub Code Generation.}
The first line of related work focuses on automatically generating stub code for mock objects.
Capture-and-replay is a popular approach adopted by these techniques.
In 2004, Saff~\etal{} were among the first to develop a mock object construction technique~\cite{DBLP:conf/paste/SaffE04,DBLP:conf/kbse/SaffAPE05} aiming to improve the efficiency of unit testing.
The technique runs a working test case to capture the interactions between the CUT and its dependencies.
Next, those dependencies are replaced with mock objects, and stub code are generated using the captured information.
A similar idea is used by Joshi~\etal{}~\cite{DBLP:conf/icsm/JoshiO07} and Elbaum~\etal{}~\cite{DBLP:conf/sigsoft/ElbaumCDD06} for test craving.
Fazzini~\etal{} proposed \textsc{MOKA}~\cite{DBLP:conf/kbse/FazziniGO20} to collect and generate mock objects for testing mobile applications by observing the interactions between the application and its environment.
More recently, Tiwari \etal{} designed \textsc{Rick}~\cite{DBLP:journals/corr/abs-2208-01321} to generate mock objects that mimic the behavior of the test dependencies in a production environment.
\textsc{Rick} works by analyzing the runtime data captured in production systems and it successfully mimics 52.4\% of the test executions as shown in evaluation.

These capture-and-replay techniques assume that dependencies are available when stub code is created.
Conversely, \tool{} does not make this assumption as it generates stub calls without executing the actual dependency, which makes it applicable to a wider range of scenarios \eg{for projects adopting TDD, the test dependencies may not be available when the test case is created}.
Moreover, capture-and-replay techniques may generate unreliable test cases when the captured behavior of the dependency is flaky or incorrect. Differently, \tool{} does not suffer from this issue.

Stub code is also generated by a few test generation techniques to increase test coverage.
For instance, Arcuri~\etal{} developed techniques for generating stub code for environment dependent classes~\cite{DBLP:conf/kbse/ArcuriFG14,DBLP:conf/sigsoft/ArcuriFG15}, which enable \textsc{EvoSuite}~\cite{DBLP:conf/sigsoft/FraserA11} to achieve higher coverage for the classes having such dependencies.
Similar approaches are also adopted to construct stub code for databases~\cite{DBLP:conf/kbse/TanejaZX10}, mobile apps~\cite{DBLP:conf/kbse/FazziniGO20}, and web services~\cite{DBLP:conf/apsec/Bhagya0G19,DBLP:journals/software/ZhangMLXTH12}.
However, they can  generate stub code for certain dependency types only (e.g., networking~\cite{DBLP:conf/sigsoft/ArcuriFG15,DBLP:conf/apsec/Bhagya0G19,DBLP:journals/software/ZhangMLXTH12}, database~\cite{DBLP:conf/kbse/TanejaZX10}, file system~\cite{DBLP:conf/kbse/ArcuriFG14,DBLP:conf/icse/MarriXTHS09}) because they follow predefined rules, which are not applicable to an arbitrary mock object.

These techniques are closely coupled with domain knowledge. New rules have to be manually defined to generate stub code for the mock objects that are not considered by these techniques.
As such, these techniques cannot be easily adapted to other types of mock objects.
Also, they do not use mocking frameworks to specify the behavior of the dependencies, but light-weight implementations similar to ``fake'' mock objects.
In comparison, \tool{} is domain agnostic and thus can synthesize stub code for an arbitrary mock object.
Also, it allows developers to specify the behavior of mock objects with oracle assertions, which gives developers more control over the behavior of the synthesized stub code.

\paragraph{Empirical Studies on Mocking.}
The second line of related work focuses on the practices adopted by developers when using mock objects in their projects.
Marri \etal~\cite{DBLP:conf/icse/MarriXTHS09} conducted an analysis on the usage of mock objects in testing file-system-dependent software and showed that mock objects can ease the process of unit testing.
Mostafa \etal~\cite{DBLP:conf/qsic/MostafaW14} analyzed the usage of mocking frameworks in 5,000 \textsc{Java} projects and revealed that mock objects are widely used although they are only used to substitute certain types of test dependencies.
They also raise the need for an automated technique for synthesizing stub code.
Spadini \etal~\cite{DBLP:journals/ese/SpadiniABB19} studied the usage of mock objects in three open-source projects and one commercial project.
They highlighted the practice adopted by developers when making mocking decisions found that developers choose to substitute the classes that are hard to setup with mock objects.
In addition, they reveal that stub code are frequently coupled with production code and need to be frequently updated, which make creating and maintaining stub code challenging.
More recently, Zhu \etal~\cite{DBLP:conf/kbse/ZhuWWLCSZ20} conducted an empirical study on four open-source projects and distilled 10 code-level rules that can affect the mocking decisions, based on which they proposed a machine learning based technique that recommends mocking decisions for developers.
Wang \etal{}~\cite{DBLP:conf/sigsoft/WangXYWW21} proposed an auto refactoring tool to migrate inheritance based mock objects to mocking frameworks.

All of these studies provide evidence of the popularity and importance of mocking.
They also discuss the challenges faced by developers when using mock objects.
In this paper, we propose \tool{} to automatically generate and repair stub code for mock objects, helping developers to address some of the challenges.

\paragraph{Test Case Repair.}
The third line of related work aim to repair the broken test cases due to the evolution of production code.
For example, Daniel \etal{}~\cite{daniel2009reassert} proposed \textsc{ReAssert}, a test case repair technique implemented for \junit{}.
\textsc{ReAssert} suggests repairs to failing tests to make them pass again.
The fixes suggested by it include replacing literals values and assertions.
Daniel \etal{}~\cite{daniel2010test} later enhanced the capability of \textsc{ReAssert} by proposing \textsc{Symbolic Test Repair}, which employs symbolic execution and constraint solving to update the expected values of the assertions.
Compared with \textsc{ReAssert}, \textsc{Symbolic Test Repair} can repair the test cases with complex control flow or operations on the expected values.
Similarly, Mirzaaghaei \etal{}~\cite{mirzaaghaei2012supporting} developed \textsc{TestCareAssistant} (TCA) to facilitate test evolution by repairing obsolete test cases and generating new test cases.
TCA identify five common actions adopted by developers to adapt the test cases to new version, and apply these actions to the obsolete test case.
While the above techniques can make the test pass again, they did not consider whether the intent of the test case are preserved.
To fill this gap, Li \etal{}~\cite{li2019intent} proposed a technique for preserving the intent of the test case during test repair.
They rank the repair candidates by the likelihood of preserving the intent of the original test case.
The intent of a test case is characterized by analyzing the path conditions generated from a dynamic symbolic execution.

Test case repair techniques are also developed for GUI or web applications.
Choudhary \etal{}~\cite{choudhary2011water} proposed \textsc{WATeR} to suggest repairs for automation script for testing web applications.
The repairs are suggest by analyzing the the difference between a passing-failing pairs.
\textsc{WATeR} can suggest repairs for the test failure due to the type change of the web page elements and displaced or changed web page elements.
Similarly, Stocco \etal{}~\cite{stocco2018visual} proposed \textsc{Vista} to repair the test script of web applications by analyzing the visual information captured from test execution.
They also equipped \textsc{Vista} with a local crawling mechanism to handle non-trivial breakage scenarios.
On the same theme, Gao \etal{}~\cite{gao2015sitar} developed \textsc{SITAR}, a semi-automated technique for repairing GUI test scripts.
The repair is generated by reverse engineering the test script and map it to an event-flow graph.
\textsc{SITAR} can amortizes the cost of human intervention across repairing multiple test scripts.

Although these techniques can effectively repair broken test cases, they focus on fixing the test exercise sequence and the assertions.
They are not capable for repairing the obsolete stub code in broken test cases.
In this paper, we proposed an application scenario for repairing the broken test cases by re-synthesizing the stub code to replace the obsolete ones.

\section{Conclusions and Future Work}\label{sec:conclusion}
Mocking is an essential part of unit testing, as it allows testing a CUT in isolation from its dependencies~\cite{DBLP:conf/kbse/ZhuWWLCSZ20,DBLP:journals/ese/SpadiniABB19}.
Mocking frameworks allow developers to write stub code to specify the behaviors of test dependencies when a test case invokes the CUT.
However, developing and maintaining stub code is a labor-intensive and error-prone activity~\cite{DBLP:journals/ese/SpadiniABB19}.

In this paper, we present \tool{} to automatically generate and repair stub code for regression tests.
\tool{} is based on the intuition that the feedback given by the runtime behavior of a test case can drive the synthesis of stub code.
In particular, \tool{} implements an evolutionary algorithm guided by a fitness function that measures how close a candidate stub code is to pass the test.

Our evaluation on \numOfBenchmarkEntries{} test cases from \numOfBenchmarkProjects{} open-source projects shows that \tool{} effectively synthesizes stub code.
Moreover, \tool{} outperforms its unguided variant, demonstrating the usefulness of the fitness function to steer the search towards generating test-passing stub code.
Also, our results show that \tool{} synthesizes stub code with similar behaviors as those written by developers.

To the best of our knowledge, \tool{} is the first technique of its kind.
There are several possible future work in this area.
We point out the two most promising ones.

First, a possible future work to improve \tool's effectiveness is to mine existing stub code in GitHub to learn recurrent patterns of stub calls.
Indeed, different software projects often share the same libraries as test dependencies.
Although stub code is test case specific, such recurrent patterns might help explore the search space more efficiently.
For instance, the mutation operators of \tool{} could give a higher probability to those mutations that match one of the mined recurrent patterns.

Second, some automated test generation techniques rely on mock objects to increase test coverage~\cite{DBLP:conf/icst/ArcuriFJ17,DBLP:conf/dagstuhl/AlshahwanJLFST10}.
However, such techniques do not explore the possible behaviors of mock objects during test generation.
This is because they models each mock object and their stub calls as a single mutation unit.
In this case, they cannot mutate each of the stub calls separately.
In comparison, \tool{} models the behavior of the stub code at a finer-grained level: it models each of the stub call as a mutation unit, and thus can explore more possible behaviors of mock objects.

The integration of such techniques and \tool{} will enable finer control on the behavior of mock objects and thus achieve higher test coverage.
 
\begin{acks}
	We would like to thank the anonymous reviewers for their insightful comments and suggestions.
    This work is supported by the National Science Foundation of China (Grant No. \code{61932021}),
    the Hong Kong Research Grant Council/General Research Fund (Grant No. \code{16205821}),
    the Hong Kong Research Grant Council/Research Impact Fund (Grant No. \code{R5034-18}),
    the Natural Sciences and Engineering Research Council of Canada Discovery Grant
    (Grant No. \code{RGPIN-2022-03744} and Grant No. \code{DGECR-2022-00378}),
    and the WeBank-HKUST Joint Laboratory.
\end{acks}

\bibliographystyle{ACM-Reference-Format}
\bibliography{bibliography/references,bibliography/links}


\begin{thebibliography}{59}


\ifx \showCODEN    \undefined \def \showCODEN     #1{\unskip}     \fi
\ifx \showDOI      \undefined \def \showDOI       #1{#1}\fi
\ifx \showISBNx    \undefined \def \showISBNx     #1{\unskip}     \fi
\ifx \showISBNxiii \undefined \def \showISBNxiii  #1{\unskip}     \fi
\ifx \showISSN     \undefined \def \showISSN      #1{\unskip}     \fi
\ifx \showLCCN     \undefined \def \showLCCN      #1{\unskip}     \fi
\ifx \shownote     \undefined \def \shownote      #1{#1}          \fi
\ifx \showarticletitle \undefined \def \showarticletitle #1{#1}   \fi
\ifx \showURL      \undefined \def \showURL       {\relax}        \fi
\providecommand\bibfield[2]{#2}
\providecommand\bibinfo[2]{#2}
\providecommand\natexlab[1]{#1}
\providecommand\showeprint[2][]{arXiv:#2}

\bibitem[Alshahwan et~al\mbox{.}(2010)]%
        {DBLP:conf/dagstuhl/AlshahwanJLFST10}
\bibfield{author}{\bibinfo{person}{Nadia Alshahwan}, \bibinfo{person}{Yue Jia},
  \bibinfo{person}{Kiran Lakhotia}, \bibinfo{person}{Gordon Fraser},
  \bibinfo{person}{David Shuler}, {and} \bibinfo{person}{Paolo Tonella}.}
  \bibinfo{year}{2010}\natexlab{}.
\newblock \showarticletitle{{AUTOMOCK:} Automated Synthesis of a Mock
  Environment for Test Case Generation}. In \bibinfo{booktitle}{\emph{Practical
  Software Testing: Tool Automation and Human Factors, 14.03. - 19.03.2010}}
  \emph{(\bibinfo{series}{Dagstuhl Seminar Proceedings},
  Vol.~\bibinfo{volume}{10111})}. \bibinfo{publisher}{Schloss Dagstuhl -
  Leibniz-Zentrum f{\"{u}}r Informatik, Germany}.
\newblock
\urldef\tempurl%
\url{http://drops.dagstuhl.de/opus/volltexte/2010/2618/}
\showURL{%
\tempurl}


\bibitem[Ammann and Offutt(2008)]%
        {DBLP:books/daglib/0020331}
\bibfield{author}{\bibinfo{person}{Paul Ammann} {and} \bibinfo{person}{Jeff
  Offutt}.} \bibinfo{year}{2008}\natexlab{}.
\newblock \bibinfo{booktitle}{\emph{Introduction to Software Testing}}.
\newblock \bibinfo{publisher}{Cambridge University Press}.
\newblock
\showISBNx{978-0-521-88038-1}
\urldef\tempurl%
\url{https://doi.org/10.1017/CBO9780511809163}
\showDOI{\tempurl}


\bibitem[Arcuri et~al\mbox{.}(2014)]%
        {DBLP:conf/kbse/ArcuriFG14}
\bibfield{author}{\bibinfo{person}{Andrea Arcuri}, \bibinfo{person}{Gordon
  Fraser}, {and} \bibinfo{person}{Juan~Pablo Galeotti}.}
  \bibinfo{year}{2014}\natexlab{}.
\newblock \showarticletitle{Automated unit test generation for classes with
  environment dependencies}. In \bibinfo{booktitle}{\emph{{ACM/IEEE}
  International Conference on Automated Software Engineering, {ASE} '14}}.
  \bibinfo{publisher}{{ACM}}, \bibinfo{pages}{79--90}.
\newblock
\urldef\tempurl%
\url{https://doi.org/10.1145/2642937.2642986}
\showDOI{\tempurl}


\bibitem[Arcuri et~al\mbox{.}(2015)]%
        {DBLP:conf/sigsoft/ArcuriFG15}
\bibfield{author}{\bibinfo{person}{Andrea Arcuri}, \bibinfo{person}{Gordon
  Fraser}, {and} \bibinfo{person}{Juan~Pablo Galeotti}.}
  \bibinfo{year}{2015}\natexlab{}.
\newblock \showarticletitle{Generating {TCP/UDP} network data for automated
  unit test generation}. In \bibinfo{booktitle}{\emph{Proceedings of the 2015
  10th Joint Meeting on Foundations of Software Engineering, {ESEC/FSE} 2015,
  Bergamo, Italy, August 30 - September 4, 2015}}. \bibinfo{publisher}{{ACM}},
  \bibinfo{pages}{155--165}.
\newblock
\urldef\tempurl%
\url{https://doi.org/10.1145/2786805.2786828}
\showDOI{\tempurl}


\bibitem[Arcuri et~al\mbox{.}(2017)]%
        {DBLP:conf/icst/ArcuriFJ17}
\bibfield{author}{\bibinfo{person}{Andrea Arcuri}, \bibinfo{person}{Gordon
  Fraser}, {and} \bibinfo{person}{Ren{\'{e}} Just}.}
  \bibinfo{year}{2017}\natexlab{}.
\newblock \showarticletitle{Private {API} Access and Functional Mocking in
  Automated Unit Test Generation}. In \bibinfo{booktitle}{\emph{2017 {IEEE}
  International Conference on Software Testing, Verification and Validation,
  {ICST} 2017}}. \bibinfo{publisher}{{IEEE} Computer Society},
  \bibinfo{pages}{126--137}.
\newblock
\urldef\tempurl%
\url{https://doi.org/10.1109/ICST.2017.19}
\showDOI{\tempurl}


\bibitem[Bhagya et~al\mbox{.}(2019)]%
        {DBLP:conf/apsec/Bhagya0G19}
\bibfield{author}{\bibinfo{person}{Thilini Bhagya}, \bibinfo{person}{Jens
  Dietrich}, {and} \bibinfo{person}{Hans~W. Guesgen}.}
  \bibinfo{year}{2019}\natexlab{}.
\newblock \showarticletitle{Generating Mock Skeletons for Lightweight
  Web-Service Testing}. In \bibinfo{booktitle}{\emph{26th Asia-Pacific Software
  Engineering Conference, {APSEC} 2019}}. \bibinfo{publisher}{{IEEE}},
  \bibinfo{pages}{181--188}.
\newblock
\urldef\tempurl%
\url{https://doi.org/10.1109/APSEC48747.2019.00033}
\showDOI{\tempurl}


\bibitem[Blickle and Thiele(1996)]%
        {DBLP:journals/ec/BlickleT96}
\bibfield{author}{\bibinfo{person}{Tobias Blickle} {and}
  \bibinfo{person}{Lothar Thiele}.} \bibinfo{year}{1996}\natexlab{}.
\newblock \showarticletitle{A Comparison of Selection Schemes used in
  Evolutionary Algorithms}.
\newblock \bibinfo{journal}{\emph{Evol. Comput.}} \bibinfo{volume}{4},
  \bibinfo{number}{4} (\bibinfo{year}{1996}), \bibinfo{pages}{361--394}.
\newblock
\urldef\tempurl%
\url{https://doi.org/10.1162/evco.1996.4.4.361}
\showDOI{\tempurl}


\bibitem[Bloch(2008)]%
        {bloch2008effective}
\bibfield{author}{\bibinfo{person}{Joshua Bloch}.}
  \bibinfo{year}{2008}\natexlab{}.
\newblock \bibinfo{booktitle}{\emph{Effective java}}.
\newblock \bibinfo{publisher}{Addison-Wesley Professional}.
\newblock


\bibitem[Choudhary et~al\mbox{.}(2011)]%
        {choudhary2011water}
\bibfield{author}{\bibinfo{person}{Shauvik~Roy Choudhary}, \bibinfo{person}{Dan
  Zhao}, \bibinfo{person}{Husayn Versee}, {and} \bibinfo{person}{Alessandro
  Orso}.} \bibinfo{year}{2011}\natexlab{}.
\newblock \showarticletitle{Water: Web application test repair}. In
  \bibinfo{booktitle}{\emph{Proceedings of the First International Workshop on
  End-to-End Test Script Engineering}}. \bibinfo{pages}{24--29}.
\newblock


\bibitem[Coles et~al\mbox{.}(2016)]%
        {DBLP:conf/issta/ColesLHPV16}
\bibfield{author}{\bibinfo{person}{Henry Coles}, \bibinfo{person}{Thomas
  Laurent}, \bibinfo{person}{Christopher Henard}, \bibinfo{person}{Mike
  Papadakis}, {and} \bibinfo{person}{Anthony Ventresque}.}
  \bibinfo{year}{2016}\natexlab{}.
\newblock \showarticletitle{{PIT:} a practical mutation testing tool for Java
  (demo)}. In \bibinfo{booktitle}{\emph{Proceedings of the 25th International
  Symposium on Software Testing and Analysis, {ISSTA} 2016}}.
  \bibinfo{publisher}{{ACM}}, \bibinfo{pages}{449--452}.
\newblock
\urldef\tempurl%
\url{https://doi.org/10.1145/2931037.2948707}
\showDOI{\tempurl}


\bibitem[contributors(2022a)]%
        {Tool:easymock}
\bibfield{author}{\bibinfo{person}{EasyMock contributors}.}
  \bibinfo{year}{2022}\natexlab{a}.
\newblock \bibinfo{title}{EasyMock}.
\newblock
\newblock
\urldef\tempurl%
\url{https://easymock.org}
\showURL{%
\tempurl}


\bibitem[contributors(2022b)]%
        {Tool:mockito}
\bibfield{author}{\bibinfo{person}{Mockito contributors}.}
  \bibinfo{year}{2022}\natexlab{b}.
\newblock \bibinfo{title}{Mockito framework site}.
\newblock
\newblock
\urldef\tempurl%
\url{https://mockito.org}
\showURL{%
\tempurl}


\bibitem[contributors(2022c)]%
        {Tool:moq4}
\bibfield{author}{\bibinfo{person}{Moq contributors}.}
  \bibinfo{year}{2022}\natexlab{c}.
\newblock \bibinfo{title}{Moq4}.
\newblock
\newblock
\urldef\tempurl%
\url{https://github.com/moq/moq4}
\showURL{%
\tempurl}


\bibitem[Damerau(1964)]%
        {Damerau_Levenshtein_distance}
\bibfield{author}{\bibinfo{person}{Fred~J. Damerau}.}
  \bibinfo{year}{1964}\natexlab{}.
\newblock \showarticletitle{A Technique for Computer Detection and Correction
  of Spelling Errors}.
\newblock \bibinfo{journal}{\emph{Commun. ACM}} \bibinfo{volume}{7},
  \bibinfo{number}{3} (\bibinfo{date}{mar} \bibinfo{year}{1964}),
  \bibinfo{pages}{171–176}.
\newblock
\showISSN{0001-0782}
\urldef\tempurl%
\url{https://doi.org/10.1145/363958.363994}
\showDOI{\tempurl}


\bibitem[Daniel et~al\mbox{.}(2010)]%
        {daniel2010test}
\bibfield{author}{\bibinfo{person}{Brett Daniel}, \bibinfo{person}{Tihomir
  Gvero}, {and} \bibinfo{person}{Darko Marinov}.}
  \bibinfo{year}{2010}\natexlab{}.
\newblock \showarticletitle{On test repair using symbolic execution}. In
  \bibinfo{booktitle}{\emph{Proceedings of the 19th international symposium on
  Software testing and analysis}}. \bibinfo{pages}{207--218}.
\newblock


\bibitem[Daniel et~al\mbox{.}(2009)]%
        {daniel2009reassert}
\bibfield{author}{\bibinfo{person}{Brett Daniel}, \bibinfo{person}{Vilas
  Jagannath}, \bibinfo{person}{Danny Dig}, {and} \bibinfo{person}{Darko
  Marinov}.} \bibinfo{year}{2009}\natexlab{}.
\newblock \showarticletitle{ReAssert: Suggesting repairs for broken unit
  tests}. In \bibinfo{booktitle}{\emph{2009 IEEE/ACM International Conference
  on Automated Software Engineering}}. IEEE, \bibinfo{pages}{433--444}.
\newblock


\bibitem[Deb et~al\mbox{.}(2002)]%
        {DBLP:journals/tec/DebAPM02}
\bibfield{author}{\bibinfo{person}{Kalyanmoy Deb}, \bibinfo{person}{Samir
  Agrawal}, \bibinfo{person}{Amrit Pratap}, {and} \bibinfo{person}{T.
  Meyarivan}.} \bibinfo{year}{2002}\natexlab{}.
\newblock \showarticletitle{A fast and elitist multiobjective genetic
  algorithm: {NSGA-II}}.
\newblock \bibinfo{journal}{\emph{{IEEE} Trans. Evol. Comput.}}
  \bibinfo{volume}{6}, \bibinfo{number}{2} (\bibinfo{year}{2002}),
  \bibinfo{pages}{182--197}.
\newblock
\urldef\tempurl%
\url{https://doi.org/10.1109/4235.996017}
\showDOI{\tempurl}


\bibitem[Elbaum et~al\mbox{.}(2006)]%
        {DBLP:conf/sigsoft/ElbaumCDD06}
\bibfield{author}{\bibinfo{person}{Sebastian~G. Elbaum},
  \bibinfo{person}{Hui~Nee Chin}, \bibinfo{person}{Matthew~B. Dwyer}, {and}
  \bibinfo{person}{Jonathan Dokulil}.} \bibinfo{year}{2006}\natexlab{}.
\newblock \showarticletitle{Carving differential unit test cases from system
  test cases}. In \bibinfo{booktitle}{\emph{Proceedings of the 14th {ACM}
  {SIGSOFT} International Symposium on Foundations of Software Engineering,
  {FSE} 2006}}. \bibinfo{publisher}{{ACM}}, \bibinfo{pages}{253--264}.
\newblock
\urldef\tempurl%
\url{https://doi.org/10.1145/1181775.1181806}
\showDOI{\tempurl}


\bibitem[Falleri et~al\mbox{.}(2014)]%
        {DBLP:conf/kbse/FalleriMBMM14}
\bibfield{author}{\bibinfo{person}{Jean{-}R{\'{e}}my Falleri},
  \bibinfo{person}{Flor{\'{e}}al Morandat}, \bibinfo{person}{Xavier Blanc},
  \bibinfo{person}{Matias Martinez}, {and} \bibinfo{person}{Martin Monperrus}.}
  \bibinfo{year}{2014}\natexlab{}.
\newblock \showarticletitle{Fine-grained and accurate source code
  differencing}. In \bibinfo{booktitle}{\emph{{ACM/IEEE} International
  Conference on Automated Software Engineering, {ASE} '14}}.
  \bibinfo{publisher}{{ACM}}, \bibinfo{pages}{313--324}.
\newblock
\urldef\tempurl%
\url{https://doi.org/10.1145/2642937.2642982}
\showDOI{\tempurl}


\bibitem[Fazzini et~al\mbox{.}(2022)]%
        {DBLP:conf/icse/FazziniCCLKGO22}
\bibfield{author}{\bibinfo{person}{Mattia Fazzini}, \bibinfo{person}{Chase
  Choi}, \bibinfo{person}{Juan~Manuel Copia}, \bibinfo{person}{Gabriel Lee},
  \bibinfo{person}{Yoshiki Kakehi}, \bibinfo{person}{Alessandra Gorla}, {and}
  \bibinfo{person}{Alessandro Orso}.} \bibinfo{year}{2022}\natexlab{}.
\newblock \showarticletitle{Use of Test Doubles in Android Testing: An In-Depth
  Investigation}. In \bibinfo{booktitle}{\emph{44th {IEEE/ACM} 44th
  International Conference on Software Engineering, {ICSE} 2022, Pittsburgh,
  PA, USA, May 25-27, 2022}}. \bibinfo{publisher}{{ACM}},
  \bibinfo{pages}{2266--2278}.
\newblock
\urldef\tempurl%
\url{https://doi.org/10.1145/3510003.3510175}
\showDOI{\tempurl}


\bibitem[Fazzini et~al\mbox{.}(2020)]%
        {DBLP:conf/kbse/FazziniGO20}
\bibfield{author}{\bibinfo{person}{Mattia Fazzini}, \bibinfo{person}{Alessandra
  Gorla}, {and} \bibinfo{person}{Alessandro Orso}.}
  \bibinfo{year}{2020}\natexlab{}.
\newblock \showarticletitle{A Framework for Automated Test Mocking of Mobile
  Apps}. In \bibinfo{booktitle}{\emph{35th {IEEE/ACM} International Conference
  on Automated Software Engineering, {ASE} 2020}}. \bibinfo{publisher}{{IEEE}},
  \bibinfo{pages}{1204--1208}.
\newblock
\urldef\tempurl%
\url{https://doi.org/10.1145/3324884.3418927}
\showDOI{\tempurl}


\bibitem[Ferrante et~al\mbox{.}(1987)]%
        {DBLP:journals/toplas/FerranteOW87}
\bibfield{author}{\bibinfo{person}{Jeanne Ferrante}, \bibinfo{person}{Karl~J.
  Ottenstein}, {and} \bibinfo{person}{Joe~D. Warren}.}
  \bibinfo{year}{1987}\natexlab{}.
\newblock \showarticletitle{The Program Dependence Graph and Its Use in
  Optimization}.
\newblock \bibinfo{journal}{\emph{{ACM} Trans. Program. Lang. Syst.}}
  \bibinfo{volume}{9}, \bibinfo{number}{3} (\bibinfo{year}{1987}),
  \bibinfo{pages}{319--349}.
\newblock
\urldef\tempurl%
\url{https://doi.org/10.1145/24039.24041}
\showDOI{\tempurl}


\bibitem[Fraser and Arcuri(2011)]%
        {DBLP:conf/sigsoft/FraserA11}
\bibfield{author}{\bibinfo{person}{Gordon Fraser} {and} \bibinfo{person}{Andrea
  Arcuri}.} \bibinfo{year}{2011}\natexlab{}.
\newblock \showarticletitle{EvoSuite: automatic test suite generation for
  object-oriented software}. In \bibinfo{booktitle}{\emph{SIGSOFT/FSE'11 19th
  {ACM} {SIGSOFT} Symposium on the Foundations of Software Engineering
  {(FSE-19)} and ESEC'11: 13th European Software Engineering Conference
  (ESEC-13)}}. \bibinfo{publisher}{{ACM}}, \bibinfo{pages}{416--419}.
\newblock
\urldef\tempurl%
\url{https://doi.org/10.1145/2025113.2025179}
\showDOI{\tempurl}


\bibitem[Fraser and Zeller(2010)]%
        {DBLP:conf/issta/FraserZ10}
\bibfield{author}{\bibinfo{person}{Gordon Fraser} {and}
  \bibinfo{person}{Andreas Zeller}.} \bibinfo{year}{2010}\natexlab{}.
\newblock \showarticletitle{Mutation-driven generation of unit tests and
  oracles}. In \bibinfo{booktitle}{\emph{Proceedings of the Nineteenth
  International Symposium on Software Testing and Analysis, {ISSTA} 2010}}.
  \bibinfo{publisher}{{ACM}}, \bibinfo{pages}{147--158}.
\newblock
\urldef\tempurl%
\url{https://doi.org/10.1145/1831708.1831728}
\showDOI{\tempurl}


\bibitem[Fundations(2022)]%
        {Tool:apachecommons}
\bibfield{author}{\bibinfo{person}{Apache~Software Fundations}.}
  \bibinfo{year}{2022}\natexlab{}.
\newblock \bibinfo{title}{Apache Commons}.
\newblock
\newblock
\urldef\tempurl%
\url{https://commons.apache.org}
\showURL{%
\tempurl}


\bibitem[Gao et~al\mbox{.}(2015)]%
        {gao2015sitar}
\bibfield{author}{\bibinfo{person}{Zebao Gao}, \bibinfo{person}{Zhenyu Chen},
  \bibinfo{person}{Yunxiao Zou}, {and} \bibinfo{person}{Atif~M Memon}.}
  \bibinfo{year}{2015}\natexlab{}.
\newblock \showarticletitle{SITAR: GUI test script repair}.
\newblock \bibinfo{journal}{\emph{Ieee transactions on software engineering}}
  \bibinfo{volume}{42}, \bibinfo{number}{2} (\bibinfo{year}{2015}),
  \bibinfo{pages}{170--186}.
\newblock


\bibitem[GitHub(2022)]%
        {GitHub:spb/cfaftest}
\bibfield{author}{\bibinfo{person}{GitHub}.} \bibinfo{year}{2022}\natexlab{}.
\newblock \bibinfo{title}{CloudFoundryApplicationFactoryTest of Spring Boot
  Admin}.
\newblock
\newblock
\urldef\tempurl%
\url{https://github.com/codecentric/spring-boot-admin/blob/d0085edfc757e1a83eb2ad4bf8f4764d2819eb6d/spring-boot-admin-client/src/test/java/de/codecentric/boot/admin/client/registration/CloudFoundryApplicationFactoryTest.java#L52}
\showURL{%
\tempurl}


\bibitem[Gosling et~al\mbox{.}(2018)]%
        {jls11}
\bibfield{author}{\bibinfo{person}{James Gosling}, \bibinfo{person}{Bill Joy},
  \bibinfo{person}{Guy Steele}, \bibinfo{person}{Gilad Bracha},
  \bibinfo{person}{Alex Buckley}, {and} \bibinfo{person}{Daniel Smith}.}
  \bibinfo{year}{2018}\natexlab{}.
\newblock \bibinfo{booktitle}{\emph{The Java® Language Specification}
  (\bibinfo{edition}{java se 11 edition} ed.)}.
\newblock
\urldef\tempurl%
\url{https://docs.oracle.com/javase/specs/jls/se11/html/index.html}
\showURL{%
\tempurl}


\bibitem[Goues et~al\mbox{.}(2012)]%
        {DBLP:journals/tse/GouesNFW12}
\bibfield{author}{\bibinfo{person}{Claire~Le Goues}, \bibinfo{person}{ThanhVu
  Nguyen}, \bibinfo{person}{Stephanie Forrest}, {and} \bibinfo{person}{Westley
  Weimer}.} \bibinfo{year}{2012}\natexlab{}.
\newblock \showarticletitle{GenProg: {A} Generic Method for Automatic Software
  Repair}.
\newblock \bibinfo{journal}{\emph{{IEEE} Trans. Software Eng.}}
  \bibinfo{volume}{38}, \bibinfo{number}{1} (\bibinfo{year}{2012}),
  \bibinfo{pages}{54--72}.
\newblock
\urldef\tempurl%
\url{https://doi.org/10.1109/TSE.2011.104}
\showDOI{\tempurl}


\bibitem[Inc.(2022)]%
        {github}
\bibfield{author}{\bibinfo{person}{GitHub Inc.}}
  \bibinfo{year}{2022}\natexlab{}.
\newblock \bibinfo{title}{GitHub}.
\newblock
\newblock
\urldef\tempurl%
\url{https://github.com}
\showURL{%
\tempurl}


\bibitem[Jia and Harman(2011)]%
        {DBLP:journals/tse/JiaH11}
\bibfield{author}{\bibinfo{person}{Yue Jia} {and} \bibinfo{person}{Mark
  Harman}.} \bibinfo{year}{2011}\natexlab{}.
\newblock \showarticletitle{An Analysis and Survey of the Development of
  Mutation Testing}.
\newblock \bibinfo{journal}{\emph{{IEEE} Trans. Software Eng.}}
  \bibinfo{volume}{37}, \bibinfo{number}{5} (\bibinfo{year}{2011}),
  \bibinfo{pages}{649--678}.
\newblock
\urldef\tempurl%
\url{https://doi.org/10.1109/TSE.2010.62}
\showDOI{\tempurl}


\bibitem[Joshi and Orso(2007)]%
        {DBLP:conf/icsm/JoshiO07}
\bibfield{author}{\bibinfo{person}{Shrinivas Joshi} {and}
  \bibinfo{person}{Alessandro Orso}.} \bibinfo{year}{2007}\natexlab{}.
\newblock \showarticletitle{{SCARPE:} {A} Technique and Tool for Selective
  Capture and Replay of Program Executions}. In \bibinfo{booktitle}{\emph{23rd
  {IEEE} International Conference on Software Maintenance {(ICSM} 2007)}}.
  \bibinfo{publisher}{{IEEE} Computer Society}, \bibinfo{pages}{234--243}.
\newblock
\urldef\tempurl%
\url{https://doi.org/10.1109/ICSM.2007.4362636}
\showDOI{\tempurl}


\bibitem[Lavinas et~al\mbox{.}(2018)]%
        {DBLP:conf/smc/LavinasASL18}
\bibfield{author}{\bibinfo{person}{Yuri~Cossich Lavinas},
  \bibinfo{person}{Claus Aranha}, \bibinfo{person}{Tetsuya Sakurai}, {and}
  \bibinfo{person}{Marcelo Ladeira}.} \bibinfo{year}{2018}\natexlab{}.
\newblock \showarticletitle{Experimental Analysis of the Tournament Size on
  Genetic Algorithms}. In \bibinfo{booktitle}{\emph{{IEEE} International
  Conference on Systems, Man, and Cybernetics, {SMC} 2018, Miyazaki, Japan,
  October 7-10, 2018}}. \bibinfo{publisher}{{IEEE}},
  \bibinfo{pages}{3647--3653}.
\newblock
\urldef\tempurl%
\url{https://doi.org/10.1109/SMC.2018.00617}
\showDOI{\tempurl}


\bibitem[Legg et~al\mbox{.}(2004)]%
        {DBLP:conf/cec/LeggHK04}
\bibfield{author}{\bibinfo{person}{Shane Legg}, \bibinfo{person}{Marcus
  Hutter}, {and} \bibinfo{person}{Akshat Kumar}.}
  \bibinfo{year}{2004}\natexlab{}.
\newblock \showarticletitle{Tournament versus fitness uniform selection}. In
  \bibinfo{booktitle}{\emph{Proceedings of the {IEEE} Congress on Evolutionary
  Computation, {CEC} 2004}}. \bibinfo{publisher}{{IEEE}},
  \bibinfo{pages}{2144--2151}.
\newblock
\urldef\tempurl%
\url{https://doi.org/10.1109/CEC.2004.1331162}
\showDOI{\tempurl}


\bibitem[Li et~al\mbox{.}(2019)]%
        {li2019intent}
\bibfield{author}{\bibinfo{person}{Xiangyu Li}, \bibinfo{person}{Marcelo
  d'Amorim}, {and} \bibinfo{person}{Alessandro Orso}.}
  \bibinfo{year}{2019}\natexlab{}.
\newblock \showarticletitle{Intent-preserving test repair}. In
  \bibinfo{booktitle}{\emph{2019 12th IEEE Conference on Software Testing,
  Validation and Verification (ICST)}}. IEEE, \bibinfo{pages}{217--227}.
\newblock


\bibitem[Marri et~al\mbox{.}(2009)]%
        {DBLP:conf/icse/MarriXTHS09}
\bibfield{author}{\bibinfo{person}{Madhuri~R. Marri}, \bibinfo{person}{Tao
  Xie}, \bibinfo{person}{Nikolai Tillmann}, \bibinfo{person}{Jonathan de
  Halleux}, {and} \bibinfo{person}{Wolfram Schulte}.}
  \bibinfo{year}{2009}\natexlab{}.
\newblock \showarticletitle{An Empirical Study of Testing File-System-Dependent
  Software with Mock Objects}. In \bibinfo{booktitle}{\emph{Proceedings of the
  4th International Workshop on Automation of Software Test, {AST} 2009}}.
  \bibinfo{publisher}{{IEEE} Computer Society}, \bibinfo{pages}{149--153}.
\newblock
\urldef\tempurl%
\url{https://doi.org/10.1109/IWAST.2009.5069054}
\showDOI{\tempurl}


\bibitem[Martinez and Monperrus(2018)]%
        {DBLP:conf/ssbse/MartinezM18}
\bibfield{author}{\bibinfo{person}{Matias Martinez} {and}
  \bibinfo{person}{Martin Monperrus}.} \bibinfo{year}{2018}\natexlab{}.
\newblock \showarticletitle{Ultra-Large Repair Search Space with Automatically
  Mined Templates: The Cardumen Mode of Astor}. In
  \bibinfo{booktitle}{\emph{Search-Based Software Engineering - 10th
  International Symposium, {SSBSE} 2018, Montpellier, France, September 8-9,
  2018, Proceedings}} \emph{(\bibinfo{series}{Lecture Notes in Computer
  Science}, Vol.~\bibinfo{volume}{11036})},
  \bibfield{editor}{\bibinfo{person}{Thelma~Elita Colanzi} {and}
  \bibinfo{person}{Phil McMinn}} (Eds.). \bibinfo{publisher}{Springer},
  \bibinfo{pages}{65--86}.
\newblock
\urldef\tempurl%
\url{https://doi.org/10.1007/978-3-319-99241-9\_3}
\showDOI{\tempurl}


\bibitem[McDonough(2021)]%
        {mcdonough2021test}
\bibfield{author}{\bibinfo{person}{James~E McDonough}.}
  \bibinfo{year}{2021}\natexlab{}.
\newblock \showarticletitle{Test Doubles}.
\newblock In \bibinfo{booktitle}{\emph{Automated Unit Testing with ABAP}}.
  \bibinfo{publisher}{Springer}, \bibinfo{pages}{159--210}.
\newblock


\bibitem[Miller and Goldberg(1995)]%
        {DBLP:journals/compsys/MillerG95}
\bibfield{author}{\bibinfo{person}{Brad~L. Miller} {and}
  \bibinfo{person}{David~E. Goldberg}.} \bibinfo{year}{1995}\natexlab{}.
\newblock \showarticletitle{Genetic Algorithms, Tournament Selection, and the
  Effects of Noise}.
\newblock \bibinfo{journal}{\emph{Complex Syst.}} \bibinfo{volume}{9},
  \bibinfo{number}{3} (\bibinfo{year}{1995}).
\newblock
\urldef\tempurl%
\url{http://www.complex-systems.com/abstracts/v09\_i03\_a02.html}
\showURL{%
\tempurl}


\bibitem[Mirzaaghaei et~al\mbox{.}(2012)]%
        {mirzaaghaei2012supporting}
\bibfield{author}{\bibinfo{person}{Mehdi Mirzaaghaei},
  \bibinfo{person}{Fabrizio Pastore}, {and} \bibinfo{person}{Mauro Pezz{\`e}}.}
  \bibinfo{year}{2012}\natexlab{}.
\newblock \showarticletitle{Supporting test suite evolution through test case
  adaptation}. In \bibinfo{booktitle}{\emph{2012 IEEE Fifth International
  Conference on Software Testing, Verification and Validation}}. IEEE,
  \bibinfo{pages}{231--240}.
\newblock


\bibitem[Mostafa and Wang(2014)]%
        {DBLP:conf/qsic/MostafaW14}
\bibfield{author}{\bibinfo{person}{Shaikh Mostafa} {and}
  \bibinfo{person}{Xiaoyin Wang}.} \bibinfo{year}{2014}\natexlab{}.
\newblock \showarticletitle{An Empirical Study on the Usage of Mocking
  Frameworks in Software Testing}. In \bibinfo{booktitle}{\emph{2014 14th
  International Conference on Quality Software}}. \bibinfo{publisher}{{IEEE}},
  \bibinfo{pages}{127--132}.
\newblock
\urldef\tempurl%
\url{https://doi.org/10.1109/QSIC.2014.19}
\showDOI{\tempurl}


\bibitem[Murphy(1996)]%
        {jaccard}
\bibfield{author}{\bibinfo{person}{Allan~H. Murphy}.}
  \bibinfo{year}{1996}\natexlab{}.
\newblock \showarticletitle{The Finley Affair: A Signal Event in the History of
  Forecast Verification}.
\newblock \bibinfo{journal}{\emph{Weather and Forecasting}}
  \bibinfo{volume}{11}, \bibinfo{number}{1} (\bibinfo{year}{1996}),
  \bibinfo{pages}{3 -- 20}.
\newblock
\urldef\tempurl%
\url{https://doi.org/10.1175/1520-0434(1996)011<0003:TFAASE>2.0.CO;2}
\showDOI{\tempurl}


\bibitem[Navarro(2001)]%
        {DBLP:journals/csur/Navarro01}
\bibfield{author}{\bibinfo{person}{Gonzalo Navarro}.}
  \bibinfo{year}{2001}\natexlab{}.
\newblock \showarticletitle{A guided tour to approximate string matching}.
\newblock \bibinfo{journal}{\emph{{ACM} Comput. Surv.}} \bibinfo{volume}{33},
  \bibinfo{number}{1} (\bibinfo{year}{2001}), \bibinfo{pages}{31--88}.
\newblock
\urldef\tempurl%
\url{https://doi.org/10.1145/375360.375365}
\showDOI{\tempurl}


\bibitem[Panichella et~al\mbox{.}(2015)]%
        {DBLP:journals/tse/PanichellaOPL15}
\bibfield{author}{\bibinfo{person}{Annibale Panichella}, \bibinfo{person}{Rocco
  Oliveto}, \bibinfo{person}{Massimiliano~Di Penta}, {and}
  \bibinfo{person}{Andrea~De Lucia}.} \bibinfo{year}{2015}\natexlab{}.
\newblock \showarticletitle{Improving Multi-Objective Test Case Selection by
  Injecting Diversity in Genetic Algorithms}.
\newblock \bibinfo{journal}{\emph{{IEEE} Trans. Software Eng.}}
  \bibinfo{volume}{41}, \bibinfo{number}{4} (\bibinfo{year}{2015}),
  \bibinfo{pages}{358--383}.
\newblock
\urldef\tempurl%
\url{https://doi.org/10.1109/TSE.2014.2364175}
\showDOI{\tempurl}


\bibitem[Saff et~al\mbox{.}(2005)]%
        {DBLP:conf/kbse/SaffAPE05}
\bibfield{author}{\bibinfo{person}{David Saff}, \bibinfo{person}{Shay Artzi},
  \bibinfo{person}{Jeff~H. Perkins}, {and} \bibinfo{person}{Michael~D. Ernst}.}
  \bibinfo{year}{2005}\natexlab{}.
\newblock \showarticletitle{Automatic test factoring for java}. In
  \bibinfo{booktitle}{\emph{20th {IEEE/ACM} International Conference on
  Automated Software Engineering {(ASE} 2005)}}. \bibinfo{publisher}{{ACM}},
  \bibinfo{pages}{114--123}.
\newblock
\urldef\tempurl%
\url{https://doi.org/10.1145/1101908.1101927}
\showDOI{\tempurl}


\bibitem[Saff and Ernst(2004)]%
        {DBLP:conf/paste/SaffE04}
\bibfield{author}{\bibinfo{person}{David Saff} {and}
  \bibinfo{person}{Michael~D. Ernst}.} \bibinfo{year}{2004}\natexlab{}.
\newblock \showarticletitle{Mock object creation for test factoring}. In
  \bibinfo{booktitle}{\emph{Proceedings of the 2004 {ACM} {SIGPLAN-SIGSOFT}
  Workshop on Program Analysis For Software Tools and Engineering, PASTE'04}}.
  \bibinfo{publisher}{{ACM}}, \bibinfo{pages}{49--51}.
\newblock
\urldef\tempurl%
\url{https://doi.org/10.1145/996821.996838}
\showDOI{\tempurl}


\bibitem[Spadini et~al\mbox{.}(2019)]%
        {DBLP:journals/ese/SpadiniABB19}
\bibfield{author}{\bibinfo{person}{Davide Spadini},
  \bibinfo{person}{Mauricio~Finavaro Aniche}, \bibinfo{person}{Magiel
  Bruntink}, {and} \bibinfo{person}{Alberto Bacchelli}.}
  \bibinfo{year}{2019}\natexlab{}.
\newblock \showarticletitle{Mock objects for testing java systems - Why and how
  developers use them, and how they evolve}.
\newblock \bibinfo{journal}{\emph{Empir. Softw. Eng.}} \bibinfo{volume}{24},
  \bibinfo{number}{3} (\bibinfo{year}{2019}), \bibinfo{pages}{1461--1498}.
\newblock
\urldef\tempurl%
\url{https://doi.org/10.1007/s10664-018-9663-0}
\showDOI{\tempurl}


\bibitem[Stocco et~al\mbox{.}(2018)]%
        {stocco2018visual}
\bibfield{author}{\bibinfo{person}{Andrea Stocco},
  \bibinfo{person}{Rahulkrishna Yandrapally}, {and} \bibinfo{person}{Ali
  Mesbah}.} \bibinfo{year}{2018}\natexlab{}.
\newblock \showarticletitle{Visual web test repair}. In
  \bibinfo{booktitle}{\emph{Proceedings of the 2018 26th ACM Joint Meeting on
  European Software Engineering Conference and Symposium on the Foundations of
  Software Engineering}}. \bibinfo{pages}{503--514}.
\newblock


\bibitem[Tamaki et~al\mbox{.}(1996)]%
        {DBLP:conf/icec/TamakiKK96}
\bibfield{author}{\bibinfo{person}{Hisashi Tamaki}, \bibinfo{person}{Hajime
  Kita}, {and} \bibinfo{person}{Shigenobu Kobayashi}.}
  \bibinfo{year}{1996}\natexlab{}.
\newblock \showarticletitle{Multi-Objective Optimization by Genetic Algorithms:
  {A} Review}. In \bibinfo{booktitle}{\emph{Proceedings of 1996 {IEEE}
  International Conference on Evolutionary Computation}}.
  \bibinfo{publisher}{{IEEE}}, \bibinfo{pages}{517--522}.
\newblock
\urldef\tempurl%
\url{https://doi.org/10.1109/ICEC.1996.542653}
\showDOI{\tempurl}


\bibitem[Taneja et~al\mbox{.}(2010)]%
        {DBLP:conf/kbse/TanejaZX10}
\bibfield{author}{\bibinfo{person}{Kunal Taneja}, \bibinfo{person}{Yi Zhang},
  {and} \bibinfo{person}{Tao Xie}.} \bibinfo{year}{2010}\natexlab{}.
\newblock \showarticletitle{{MODA:} automated test generation for database
  applications via mock objects}. In \bibinfo{booktitle}{\emph{{ASE} 2010, 25th
  {IEEE/ACM} International Conference on Automated Software Engineering}}.
  \bibinfo{publisher}{{ACM}}, \bibinfo{pages}{289--292}.
\newblock
\urldef\tempurl%
\url{https://doi.org/10.1145/1858996.1859053}
\showDOI{\tempurl}


\bibitem[Terragni et~al\mbox{.}(2020)]%
        {DBLP:conf/sigsoft/TerragniJTP20}
\bibfield{author}{\bibinfo{person}{Valerio Terragni}, \bibinfo{person}{Gunel
  Jahangirova}, \bibinfo{person}{Paolo Tonella}, {and} \bibinfo{person}{Mauro
  Pezz{\`{e}}}.} \bibinfo{year}{2020}\natexlab{}.
\newblock \showarticletitle{Evolutionary improvement of assertion oracles}. In
  \bibinfo{booktitle}{\emph{{ESEC/FSE} '20: 28th {ACM} Joint European Software
  Engineering Conference and Symposium on the Foundations of Software
  Engineering, Virtual Event, USA, November 8-13, 2020}},
  \bibfield{editor}{\bibinfo{person}{Prem Devanbu}, \bibinfo{person}{Myra~B.
  Cohen}, {and} \bibinfo{person}{Thomas Zimmermann}} (Eds.).
  \bibinfo{publisher}{{ACM}}, \bibinfo{pages}{1178--1189}.
\newblock
\urldef\tempurl%
\url{https://doi.org/10.1145/3368089.3409758}
\showDOI{\tempurl}


\bibitem[Thomas and Hunt(2002)]%
        {thomas2002mock}
\bibfield{author}{\bibinfo{person}{Dave Thomas} {and} \bibinfo{person}{Andy
  Hunt}.} \bibinfo{year}{2002}\natexlab{}.
\newblock \showarticletitle{Mock objects}.
\newblock \bibinfo{journal}{\emph{IEEE Software}} \bibinfo{volume}{19},
  \bibinfo{number}{3} (\bibinfo{year}{2002}), \bibinfo{pages}{22--24}.
\newblock


\bibitem[Tiwari et~al\mbox{.}(2022)]%
        {DBLP:journals/corr/abs-2208-01321}
\bibfield{author}{\bibinfo{person}{Deepika Tiwari}, \bibinfo{person}{Martin
  Monperrus}, {and} \bibinfo{person}{Benoit Baudry}.}
  \bibinfo{year}{2022}\natexlab{}.
\newblock \showarticletitle{Mimicking Production Behavior with Generated
  Mocks}.
\newblock \bibinfo{journal}{\emph{CoRR}}  \bibinfo{volume}{abs/2208.01321}
  (\bibinfo{year}{2022}).
\newblock
\urldef\tempurl%
\url{https://doi.org/10.48550/arXiv.2208.01321}
\showDOI{\tempurl}
\showeprint[arXiv]{2208.01321}


\bibitem[Wang et~al\mbox{.}(2021)]%
        {DBLP:conf/sigsoft/WangXYWW21}
\bibfield{author}{\bibinfo{person}{Xiao Wang}, \bibinfo{person}{Lu Xiao},
  \bibinfo{person}{Tingting Yu}, \bibinfo{person}{Anne Woepse}, {and}
  \bibinfo{person}{Sunny Wong}.} \bibinfo{year}{2021}\natexlab{}.
\newblock \showarticletitle{An automatic refactoring framework for replacing
  test-production inheritance by mocking mechanism}. In
  \bibinfo{booktitle}{\emph{{ESEC/FSE} '21: 29th {ACM} Joint European Software
  Engineering Conference and Symposium on the Foundations of Software
  Engineering}}. \bibinfo{publisher}{{ACM}}, \bibinfo{pages}{540--552}.
\newblock
\urldef\tempurl%
\url{https://doi.org/10.1145/3468264.3468590}
\showDOI{\tempurl}


\bibitem[Wegener et~al\mbox{.}(2001)]%
        {DBLP:journals/infsof/WegenerBS01}
\bibfield{author}{\bibinfo{person}{Joachim Wegener},
  \bibinfo{person}{Andr{\'{e}} Baresel}, {and} \bibinfo{person}{Harmen
  Sthamer}.} \bibinfo{year}{2001}\natexlab{}.
\newblock \showarticletitle{Evolutionary test environment for automatic
  structural testing}.
\newblock \bibinfo{journal}{\emph{Inf. Softw. Technol.}} \bibinfo{volume}{43},
  \bibinfo{number}{14} (\bibinfo{year}{2001}), \bibinfo{pages}{841--854}.
\newblock
\urldef\tempurl%
\url{https://doi.org/10.1016/S0950-5849(01)00190-2}
\showDOI{\tempurl}


\bibitem[Yu et~al\mbox{.}(2019)]%
        {DBLP:conf/icsm/YuTA19}
\bibfield{author}{\bibinfo{person}{Chak~Shun Yu}, \bibinfo{person}{Christoph
  Treude}, {and} \bibinfo{person}{Mauricio~Finavaro Aniche}.}
  \bibinfo{year}{2019}\natexlab{}.
\newblock \showarticletitle{Comprehending Test Code: An Empirical Study}. In
  \bibinfo{booktitle}{\emph{2019 {IEEE} International Conference on Software
  Maintenance and Evolution, {ICSME} 2019}}. \bibinfo{publisher}{{IEEE}},
  \bibinfo{pages}{501--512}.
\newblock
\urldef\tempurl%
\url{https://doi.org/10.1109/ICSME.2019.00084}
\showDOI{\tempurl}


\bibitem[Yuan and Banzhaf(2020)]%
        {DBLP:journals/tse/YuanB20}
\bibfield{author}{\bibinfo{person}{Yuan Yuan} {and} \bibinfo{person}{Wolfgang
  Banzhaf}.} \bibinfo{year}{2020}\natexlab{}.
\newblock \showarticletitle{{ARJA:} Automated Repair of Java Programs via
  Multi-Objective Genetic Programming}.
\newblock \bibinfo{journal}{\emph{{IEEE} Trans. Software Eng.}}
  \bibinfo{volume}{46}, \bibinfo{number}{10} (\bibinfo{year}{2020}),
  \bibinfo{pages}{1040--1067}.
\newblock
\urldef\tempurl%
\url{https://doi.org/10.1109/TSE.2018.2874648}
\showDOI{\tempurl}


\bibitem[Zhang et~al\mbox{.}(2012)]%
        {DBLP:journals/software/ZhangMLXTH12}
\bibfield{author}{\bibinfo{person}{Linghao Zhang}, \bibinfo{person}{Xiaoxing
  Ma}, \bibinfo{person}{Jian Lu}, \bibinfo{person}{Tao Xie},
  \bibinfo{person}{Nikolai Tillmann}, {and} \bibinfo{person}{Peli de Halleux}.}
  \bibinfo{year}{2012}\natexlab{}.
\newblock \showarticletitle{Environmental Modeling for Automated Cloud
  Application Testing}.
\newblock \bibinfo{journal}{\emph{{IEEE} Softw.}} \bibinfo{volume}{29},
  \bibinfo{number}{2} (\bibinfo{year}{2012}), \bibinfo{pages}{30--35}.
\newblock
\urldef\tempurl%
\url{https://doi.org/10.1109/MS.2011.158}
\showDOI{\tempurl}


\bibitem[Zhu et~al\mbox{.}(2020)]%
        {DBLP:conf/kbse/ZhuWWLCSZ20}
\bibfield{author}{\bibinfo{person}{Hengcheng Zhu}, \bibinfo{person}{Lili Wei},
  \bibinfo{person}{Ming Wen}, \bibinfo{person}{Yepang Liu},
  \bibinfo{person}{Shing{-}Chi Cheung}, \bibinfo{person}{Qin Sheng}, {and}
  \bibinfo{person}{Cui Zhou}.} \bibinfo{year}{2020}\natexlab{}.
\newblock \showarticletitle{MockSniffer: Characterizing and Recommending
  Mocking Decisions for Unit Tests}. In \bibinfo{booktitle}{\emph{35th
  {IEEE/ACM} International Conference on Automated Software Engineering, {ASE}
  2020}}. \bibinfo{publisher}{{IEEE}}, \bibinfo{pages}{436--447}.
\newblock
\urldef\tempurl%
\url{https://doi.org/10.1145/3324884.3416539}
\showDOI{\tempurl}


\end{thebibliography}

\end{document}